\documentclass[ floatfix, tightenlines, aps, prd, twocolumn, lengthcheck, sort&compress, showpacs,  nofootinbib, superscriptaddress]{revtex4-1}

\usepackage{amsmath}
\usepackage{graphicx}
\usepackage[usenames]{color}
\usepackage{dsfont}
\usepackage{slashed}
\usepackage[utf8]{inputenc}

\usepackage[colorlinks=true,linkcolor=blue,citecolor=blue,urlcolor=blue]{hyperref}

\newcommand{\ii}{\mathrm{i}}

\begin{document}

\title{Light mesons in QCD and unquenching effects from the 3PI effective action}

\author{Richard Williams}
\email{richard.williams@physik.uni-giessen.de}
\affiliation{Institut f\"ur Theoretische Physik, Justus-Liebig--Universit\"at Giessen, 35392 Giessen, Germany.}

\author{Christian S. Fischer}
\email{christian.fischer@physik.uni-giessen.de}
\affiliation{Institut f\"ur Theoretische Physik, Justus-Liebig--Universit\"at Giessen, 35392 Giessen, Germany.}

\author{Walter Heupel}
\email{walter.heupel@physik.uni-giessen.de}
\affiliation{Institut f\"ur Theoretische Physik, Justus-Liebig--Universit\"at Giessen, 35392 Giessen, Germany.}

\begin{abstract}
We investigate the impact of unquenching effects on QCD Green's functions, in the form of 
quark-loop contributions to both the gluon propagator and three-gluon vertex, in a three-loop 
inspired truncation of the three-particle irreducible ($3$PI) effective action. The fully coupled 
system of Dyson-Schwinger equations for the quark-gluon-, ghost-gluon- and three-gluon vertices, 
together with the quark propagator, are solved self-consistently; our only input are the ghost and 
gluon propagators themselves that are constrained by calculations within Lattice QCD. We find that 
the two different unquenching effects have roughly equal, but opposite, impact on the quark-gluon 
vertex and quark propagator, with an overall negative impact on the latter.
By taking further derivatives of the 3PI effective action, we construct the corresponding 
quark-antiquark kernel of the Bethe-Salpeter equation for mesons. The leading component is gluon 
exchange between two fully-dressed quark-gluon vertices, thus introducing for the first time an 
obvious scalar-scalar component to the binding.   We gain access to time-like properties of 
bound-states by analytically continuing the coupled system of Dyson--Schwinger equations to the 
complex plane. We observe that the vector axial-vector splitting is in accord with experiment and 
that the lightest quark-antiquark scalar meson is above $1$~GeV in mass.
\end{abstract}


\maketitle

\section{Introduction}\label{sec:introduction}
The plethora of experimental phenomena related to the global properties and the internal structure 
of hadronic bound states and resonances are generated from the underlying non-perturbative 
interaction of quarks and gluons described by QCD. These relations can be made apparent using 
functional methods such as the framework of Dyson-Schwinger and Bethe-Salpeter equations or the 
functional renormalization group, 
see~\cite{Roberts:1994dr,Alkofer:2000wg,Fischer:2006ub,Bashir:2012fs,Berges:2000ew,Pawlowski:2005xe} for reviews.

One of the long-standing goals within the Dyson--Schwinger/Bethe--Salpeter framework is to 
establish robust truncation schemes that can be systematically applied to the calculation of 
bound-state properties. Such constructions can be approached from two different perspectives: 
bottom-up or top-down. While the former employs phenomenological input in order to construct 
models and constrain their parameters, the latter requires a robust theoretical foundation upon 
which to build~\cite{Alkofer:2000wg}. Consequently there is a rich and diverse history regarding 
truncations, ranging from schemes that adhere to computational 
prudence~\cite{Munczek:1983dx,Roberts:2011wy}, set-ups that employ symmetries and identities to constrain Green's functions beyond 
propagators~\cite{Chang:2011ei,Heupel:2014ina,Aguilar:2014lha}, to recent investigations 
wherein vertices are solved for 
explicitly~\cite{Schleifenbaum:2004id,Kellermann:2008iw,Alkofer:2008tt,Huber:2012kd,Aguilar:2013vaa,Blum:2014gna,Eichmann:2014xya,Cyrol:2014kca,Williams:2014iea,Vujinovic:2014ioa,Mitter:2014wpa,Braun:2014ata,Hopfer:PhD,Windisch:PhD}.

Obtaining a suitable description of QCD within the functional approach is relevant for several 
diverse reasons. Aside from studying intrinsic properties of Green's functions and their 
connection to confinement, one can explore properties of the fundamental quark and gluon degrees 
of freedom in-medium, thus providing a handle on the QCD phase diagram. Additionally, composite 
systems can be constructed in the form of mesons~\cite{Fischer:2014xha,Hilger:2015ora}, 
baryons~\cite{Eichmann:2009qa,SanchisAlepuz:2011jn}, glueballs~\cite{Sanchis-Alepuz:2015hma} and 
tetraquarks~\cite{Heupel:2012ua,Eichmann:2015cra}, with their mass spectra, decays and 
electromagnetic interactions explored~\cite{Maris:1999bh,Eichmann:2011vu,Sanchis-Alepuz:2013iia}. 
This of course necessitates that key symmetries are maintained -- a principle difficulty in 
constructing viable truncations -- as we shall discuss later.

Furthermore, detailed Lattice calculations of QCD in Landau gauge are by now sufficiently advanced 
that they can serve to provide auxiliary information. This not only enables one to judge the 
efficacy of existing truncations, but to provide key ingredients or missing information. Finding 
coincidence or convergence between these complementary non-perturbative approaches enables hybrid 
constructions to be developed along the lines of Refs.~\cite{Fischer:2012vc,Fischer:2014ata}, 
wherein the difficulties of one approach (such as the sign problem) can be circumvented.

In this article, we explore one of the most important ingredients in non-perturbative studies of 
QCD that couple together the gauge and matter sector: the quark-gluon vertex. However, rather than 
following the customary approach of truncating the 1PI Dyson--Schwinger equations at the level of 
vertex functions, we take here a more pragmatic (and arguably more systematic) approach by 
truncating the $n$PI effective action to a given loop order. In particular, we take the $3$PI 
effective action to three-loops, such that all two- and three-point functions are dynamical 
quantities. This is a natural step beyond the system explored in 
Ref.~\cite{Sanchis-Alepuz:2015tha,Sanchis-Alepuz:2015qra} which is analogous to a three-loop 
truncation of the $2$PI effective action.
The resulting system of equations is still extremely complex and expensive in numerical terms, in 
particular due to the two-loop structure of the resulting Dyson-Schwinger equation (DSE) for the 
gluon propagator. We therefore solve the DSEs for the ghost and gluon propagator in a separate 
truncation providing solutions that are close to corresponding (quenched and unquenched) lattice 
results. These are then used as input into the remaining 3PI equations for the primitively 
divergent three-point vertices together with the quark propagator. We solve these self-consistently
and discuss the impact of unquenching effects on the quark-gluon and three-gluon vertices. 
Finally, we apply our approach to the Bethe-Salpeter description of mesons and determine a number 
of observables consistent with the axial Ward-Takahashi identity.
In general, our top-down approach is similar in spirit to a corresponding effort within the
framework of the functional renormalization group~\cite{Mitter:2014wpa,Braun:2014ata}.

This paper is organized as follows. In section~\ref{sec:framework} we introduce in brief the $3$PI 
effective action of QCD at three-loop order and present the relevant dynamical quantities, the 
propagators and vertices, together with their equations of motion in closed form. In 
section~\ref{sec:results} we give our results for each Green's function and discuss the impact of 
quark-loops on the three-gluon vertex. We apply the formalism to the calculation of meson 
observables in a symmetry preserving truncation of the Bethe-Salpeter equation. In 
section~\ref{sec:conclusion} we conclude; some technical details are relegated to an appendix.

\section{Framework}\label{sec:framework}
\begin{figure}[!b]
\centering\includegraphics[width=0.45\textwidth]{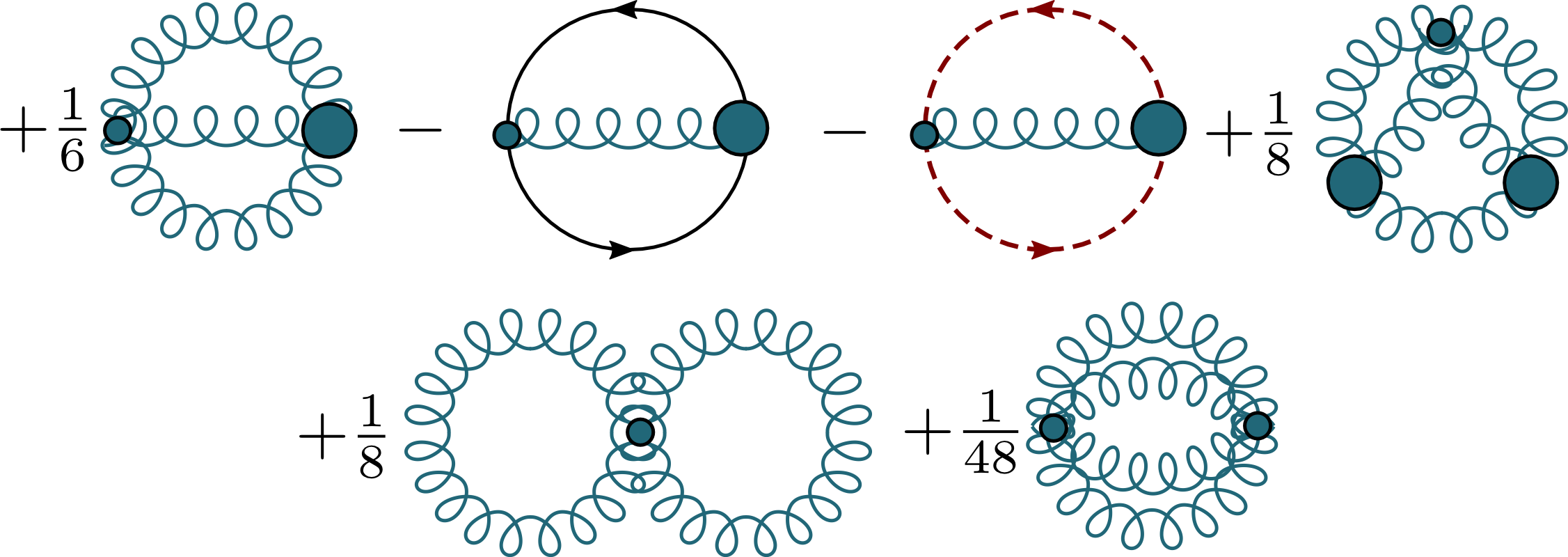}
\caption{Non-interacting part of the $3$PI effective action, to three-loop. All propagators are 
considered dressed. Throughout the paper springs describe gluons, dashed lines Faddeev-Popov 
ghosts and solid lines quarks. Small filled circles describe bare and large filled circles 
describe dressed vertices.}
\label{fig:effectiveaction0}
\end{figure}
\begin{figure}[!b]
\centering\includegraphics[width=0.45\textwidth]{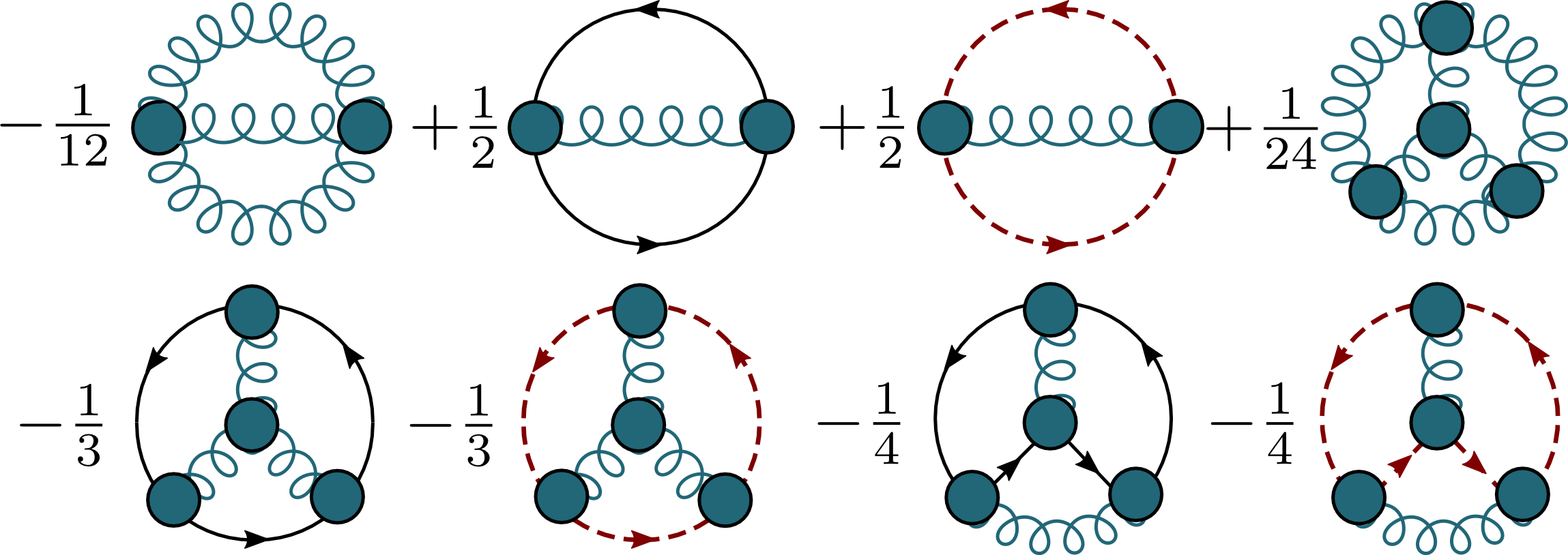}
\caption{Interacting part of the $3$PI effective action, to three-loop.}
\label{fig:effectiveactionint}
\end{figure}
In the functional approach, the Dyson--Schwinger equations are the equations of motion 
corresponding to the $1$PI effective action. They are comprised of coupled integral equations that 
form exact relations between the theory's infinite tower of $n$-point Green's functions. The 
effective action, $\Gamma\left[\phi\right]$, is obtained from the generating functional of 
connected Green's functions $W\left[J\right]$ by a Legendre transform
\begin{align}\label{eqn:action1PI}
\Gamma\left[\phi\right] = W\left[J\right] - J_i\phi_i\;,
\end{align}
from which $n$-point correlation functions are defined
\begin{align}
G_{(n)}(p_1,\ldots,p_n) = \frac{\delta^n\Gamma\left[\phi\right]}{\delta\phi_1\cdots\delta\phi_n}\;,
\end{align}
by taking functional derivatives. Then, the DSE for a $1$PI Green's function can be derived from the functional identity
\begin{align}
\frac{\delta\Gamma\left[\phi\right]}{\delta\phi_i}-\frac{\delta S}{\delta\phi_i}\left[\phi+\frac{\delta^2W[J]}{\delta J\delta J_k}\frac{\delta}{\delta\phi_k} \right]=0\;.
\end{align}
However, by themselves the DSEs do not form a closed system and require truncation. Typically this 
involves specifying the behavior of higher order $n$-point functions and collapsing the infinite 
tower to a manageable set of coupled equations.

Another approach is to work with a different resummation of the effective action by performing 
additional Legendre transformations of the action~\eqref{eqn:action1PI}, this time with respect to 
propagators and vertices~\cite{Cornwall:1974vz}. Here we consider the $3$PI effective action which 
in compact notation~\cite{Carrington:2010qq,York:2012ib,Carrington:2013koa} reads
\begin{align}\label{eqn:nPIeffectiveaction}
\Gamma\left[ \phi,D,U\right]&=S_{cl}\left[\phi\right] + \frac{i}{2}\mathrm{Tr}\mathrm{Ln}D^{-1}+\frac{i}{2}\mathrm{Tr}\left[D_{(0)}^{-1}D\right]\\
&-i\Phi^0\left[\phi,D,U\right]-i\Phi^{int}\left[\phi,D,U\right]+\mathrm{const}\;.\nonumber
\end{align}
The superfield $\phi$ represents all fields in the action, and $D$, $U$ are the corresponding 
propagators and three-point vertices. As usual, the equations of motion are obtained by taking 
functional derivatives
\begin{align}
\frac{\delta\Gamma[\phi,D,U]}{\delta D} = \frac{\delta\Gamma[\phi,D,U]}{\delta U} = 0\;.
\label{eqn:equationsofmotion}
\end{align}
The resulting set of equations for the propagators and vertices is then closed. In the case of 
QCD, the non-interacting part, $\Phi^0\left[\phi,D,U\right]$ is given in 
Fig.~\ref{fig:effectiveaction0}, and the interacting part $\Phi^{int}\left[\phi,D,U\right]$ in 
Fig.~\ref{fig:effectiveactionint}.

Note that throughout this paper we work in Euclidean space, wherein spacelike momenta are those 
for which $p^2\ge0$. To compute time-like properties of bound-states we must analytically continue 
to complex momenta; we will discuss this later in brief. For the sake of brevity we will drop the 
majority of (easily determinable) momentum arguments in the equations that follow.

\subsection{Ghost and gluon propagators}
The ghost propagator in Landau gauge is defined
\begin{align}\label{eqn:ghostpropagator}
D_G(p) = - \frac{G(p^2)}{p^2}\;,
\end{align}
with $G(p^2)$ the ghost dressing function. Similarly, the gluon propagator is
\begin{align}\label{eqn:gluonpropagator}
D^{\mu\nu}(p) = T^{\mu\nu}_{(p)}D_Z(p^2)= T^{\mu\nu}_{(p)}\frac{Z(p^2)}{p^2} \;,
\end{align}
with $Z(p^2)$ a momentum dependent dressing function. The tensor structure is given by the transverse projector
\begin{align}\label{eqn:transverseprojector}
T^{\mu\nu}_{(p)} =  \delta^{\mu\nu} - \frac{p^\mu p^\nu}{p^2}\;.
\end{align}
From \eqref{eqn:nPIeffectiveaction}--\eqref{eqn:equationsofmotion} and 
Figs.~\ref{fig:effectiveaction0},~\ref{fig:effectiveactionint} we can derive the (truncated) DSE 
for the ghost and gluon propagators, displayed in Fig.~\ref{fig:ghostgluondse} in 
appendix~\ref{sec:ghostgluon}. Unfortunately, in contrast to the other DSEs considered in this 
work, these feature two-loop diagrams with squint and sunset topology that pose a significant 
calculational challenge. This prevents a complete self-consistent solution together with the DSEs 
for the vertices\footnote{Though see~\cite{Bloch:2003yu,Meyers:2014iwa,Mader:2014qca,Hopfer:PhD} 
for some progress within simpler truncation schemes.}. On the other hand, we need quantitatively 
correct ghost and gluon propagators as input for the other DSEs we wish to solve, i.e. the one for 
the ghost-gluon vertex, the three-gluon vertex, the quark-gluon vertex and the quark.
\begin{figure}[!t]
\centering\includegraphics[width=0.5\textwidth]{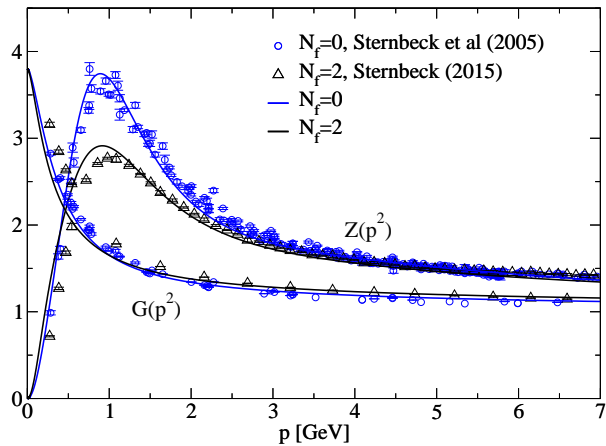}
\caption{Ghost and gluon dressing functions for the quenched ($N_f=0$) and
unquenched system ($N_f=2$) compared with lattice data from 
Refs.~\cite{Sternbeck:2005tk,Sternbeck:2015}.}
\label{fig:gluondressings}
\end{figure}

To bypass this difficulty, we employ the framework of 
Refs.~\cite{Fischer:2003rp,Fischer:2005en,Huber:2012kd} and solve a coupled system of ghost, gluon 
and (in case of $N_f \neq 0$) quark propagators using model ansaetze for the three-point vertices, 
while neglecting all diagrams that involve the four-gluon vertex (i.e. the two-loop diagrams). The 
ansaetze are chosen such that lattice data for the ghost and gluon propagators for $N_f=0$ 
(quenched) and $N_f=2$ (unquenched) quark flavors are reproduced. The advantages of this procedure 
over just using the lattice data for ghost and glue in the other DSEs are two-fold. First, we are 
able to use continuous solutions for the ghost and gluon propagators as input without the need to 
interpolate and extrapolate the lattice data. Second, we can set up and use a consistent 
renormalization scheme for all DSEs with appropriate renormalization factors $Z_3(\mu,\Lambda)$ 
and $\tilde{Z}_3(\mu,\Lambda)$ at a consistent renormalization point $\mu$ and numerical cutoff 
$\Lambda$. However, this comes at a price: neglecting the two-loop diagrams but still reproducing 
the results of the lattice calculations means that we have to use effective ghost-gluon, 
three-gluon and (to a lesser extent) quark-gluon vertex models
in the ghost-gluon DSEs
that make up for the neglected contributions. These are then no longer quantitatively consistent with the explicit vertices determined in this work.

Relegating all technical details to appendix \ref{sec:ghostgluon} we only discuss the resulting 
ghost and gluon dressing functions $G(p^2)$ and $Z(p^2)$ and compare them with corresponding 
lattice data in Fig.~\ref{fig:gluondressings}. In the figure, the data are taken from 
Ref.~\cite{Sternbeck:2005tk,Sternbeck:2015}. Similar data from other groups have been discussed in 
Ref.~\cite{Cucchieri:2007rg,Ayala:2012pb}. The agreement in the quenched case is almost perfect, 
and for the unquenched case is very good on the level of a few percent. This is certainly 
sufficient for our study.

\subsection{Quark propagator}
\begin{figure}[!h]
\centering\includegraphics[width=0.45\textwidth]{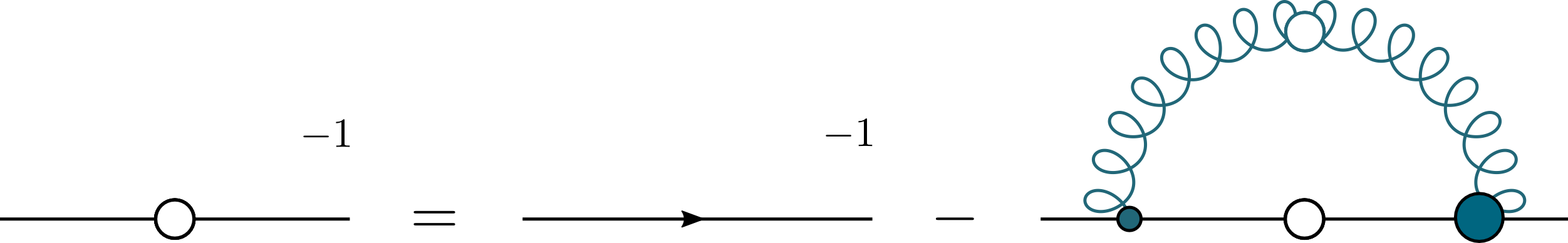}
\caption{Diagrammatic representation of the quark DSE. Blobs on propagators indicate they are 
dressed.}
\label{fig:quarkDSE}
\end{figure}

The quark propagator in vacuum has the general decomposition
\begin{align}
S^{-1}(p) = Z_f^{-1}(p^2)\left[ i \slashed{p}  + M(p^2) \right]\;,
\end{align}
where $Z_f(p^2)$ and $M(p^2)$ are the quark wave function and mass function, respectively. They 
are obtained by solving the quark gap equation, see Fig.~\ref{fig:quarkDSE}, given by
\begin{align}
S^{-1} &= Z_2 S^{-1}_{(0)}+Z_{1f}C_F g_s^2 \int_k  \gamma^\mu S \Gamma^\nu_{\mathrm{qg}} D^{\mu\nu}\;,
\end{align}
with $\int_k = d^4k/(2\pi)^4$ the integration measure, $C_F=4/3$ the result of the color trace, 
and $S^{-1}_{(0)}$ the bare inverse propagator. The quark and quark-gluon vertex renormalizations 
are $Z_2$, $Z_{1f}$ respectively. The quark-gluon vertex $\Gamma^\nu_{\mathrm{qg}}$ is detailed in 
the next subsection.

\subsection{Quark-gluon vertex}

\begin{figure}[!t]
\centering\includegraphics[width=0.45\textwidth]{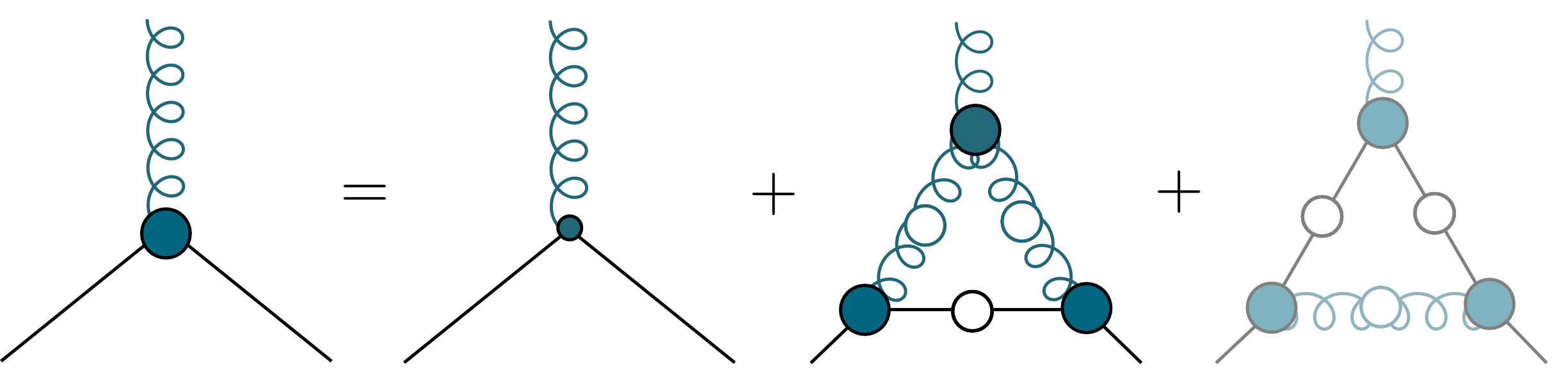}
\caption{Diagrammatic representation of the quark-gluon vertex DSE.}
\label{fig:qgvertexDSE}
\end{figure}

The quark-gluon vertex can be decomposed into a set of Dirac-Lorentz covariants $X^\mu_i$ and 
scalar dressings $h_i$
\begin{align}
\Gamma^\mu_{\mathrm{qg}}(l,k) = \sum_i h_i X^\mu_i(l,k)\;.
\end{align}
The covariants are any linear combination of the following twelve elements
$\gamma^\mu$, $\gamma^\mu \slashed{k}$, $\gamma^\mu \slashed{l}$, $\gamma^\mu \slashed{k}\slashed{l}$,
$k^\mu$, $k^\mu \slashed{k}$, $k^\mu \slashed{l}$, $k^\mu \slashed{k}\slashed{l}$,
$l^\mu$, $l^\mu \slashed{k}$, $l^\mu \slashed{l}$, $l^\mu \slashed{k}\slashed{l}$.
In Landau gauge, however, it is more convenient to work with the transversely projected vertex 
where each element transforms correctly under charge conjugation. Just eight components suffice
 \begin{align}\label{eqn:transversebasis}
T^{\mu\nu}_{(k)} \Gamma^\nu(l,k)
                             &=h_1\, \gamma^\mu_T
                             + h_2\, l^\mu_T \slashed{l}
                             + h_3\, i l^\mu_T
                             \nonumber\\
                            &+ h_4\, \left(l\cdot k\right) \frac{i}{2}\left[\gamma^\mu_T,\slashed{l}\right]
                             + h_5\, \frac{i}{2}\left[\gamma^\mu,\slashed{k}\right]
                             \nonumber\\
                            &+ h_6\, \frac{1}{6}\Big\{\left[ \gamma^\mu,\slashed{l}\right]\slashed{k} + \left[ \slashed{l},\slashed{k}\right]\gamma^\mu + \left[ \slashed{k},\gamma^\mu\right]\slashed{l}\Big\}
                            \nonumber\\
                            &+ h_7\, t^{\mu\nu}_{(kl)}\left(l\cdot k\right) \gamma^\nu
                             + h_8\, t^{\mu\nu}_{(kl)}\frac{i}{2}\left[\gamma^\nu,\slashed{l} \right]\,.
\end{align}
Here, the incoming gluon momentum is $k^\mu$, and $l^\mu$ is the relative quark momentum.  
Quantities with a subscript $T$ are contracted with the transverse projector $T^{\mu\nu}_{(k)}$, 
see \eqref{eqn:transverseprojector} and 
$\tau^{\mu\nu}_{(kl)}=\left(l\cdot k\right)\delta^{\mu\nu}-l^\mu k^\nu$.

The DSE for the vertex, following the three-loop truncation of the $3$PI effective action, is 
given in Fig.~\ref{fig:qgvertexDSE}. It consists of two vertex corrections which we refer to as 
the non-Abelian (since it involves gluon self-interaction) and Abelian diagrams
\begin{align}
\Gamma^\mu_{\mathrm{qg}} = Z_{1f}\gamma^\mu + \Lambda_{\mathrm{qg},\mathrm{NA}}^\mu + \Lambda_{\mathrm{qg},\mathrm{AB}}^\mu \;.
\end{align}
Explicitly their contributions are
\begin{align}
\Lambda_{\mathrm{qg},\mathrm{NA}}^\mu &= \frac{g_s^2N_c}{2}\int_k
\Gamma_{\mathrm{qg}}^\alpha S\Gamma_{\mathrm{qg}}^\beta\Gamma_{\mathrm{3g}}^{\alpha\beta\mu}D_Z D_Z\;,\\
\Lambda_{\mathrm{qg},\mathrm{AB}}^\mu &= \frac{-g_s^2}{2N_c}\int_k\Gamma_{\mathrm{qg}}^\alpha S\Gamma_{\mathrm{qg}}^\mu S\Gamma_{\mathrm{qg}}^{\alpha}D_Z\;.
\end{align}
Since the quark-gluon vertex is defined transverse with respect to its gluon momentum, it suffices 
to write only the scalar part of the gluon propagators, $D_Z$. Thus far the only component that 
has not been introduced is the three-gluon vertex $\Gamma_{\mathrm{3g}}^{\mu\nu\rho}$, which we 
discuss later.

When solving the Abelian diagram, it is useful to make the substitution 
$\chi_{\mathrm{qg}}^\mu = S \Gamma_{\mathrm{qg}}^\mu S$, for the top-most vertex, which 
significantly reduces the complexity of the resulting trace algebra. Despite the seeming 
complexity of the system, each iteration of the quark-gluon vertex takes about a minute on a 
standard desktop CPU.

\subsection{Ghost-gluon vertex}
\begin{figure}[!h]
\centering\includegraphics[width=0.45\textwidth]{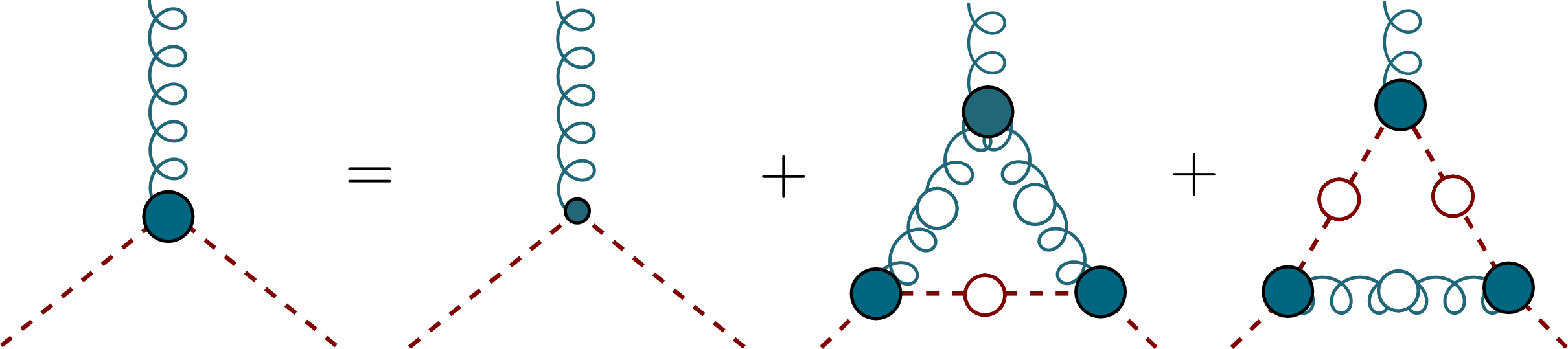}
\caption{Diagrammatic representation of the ghost-gluon vertex DSE.}
\label{fig:ghvertexDSE}
\end{figure}
For the ghost-gluon vertex in Landau gauge, there is just the one dressing function, $f(l,q)$ and 
corresponding tensor structure
\begin{align}
\Gamma^\mu_{\mathrm{gh}}(l, q) = f(l,q)\;T^{\mu\nu}_{(q)} l^\nu\,
\end{align}
with $l=(p_1 + p_2)/2$ the relative ghost momentum and $q$ the gluon momentum. The DSE, given in 
Fig.~\ref{fig:ghvertexDSE}, is similar in form to that of the quark-gluon vertex
\begin{align}
\Gamma^\mu_{\mathrm{gh}} = \tilde{Z}_{1}l^\mu_T + \Lambda_{\mathrm{gh},\mathrm{NA}}^\mu + \Lambda_{\mathrm{gh},\mathrm{AB}}^\mu \;,
\end{align}
where $ \tilde{Z}_{1}=1$ in Landau gauge. The individual contributions are given by
\begin{align}
    \Lambda_{\mathrm{gh},\mathrm{NA}}^\mu &= \frac{g_s^2N_c}{2}\int_k\Gamma_{\mathrm{gh}}^\alpha  \Gamma_{\mathrm{3g}}^{\alpha\beta\mu} \Gamma_{\mathrm{gh}}^\beta D_Z D_Z D_G\;,\\
    \Lambda_{\mathrm{gh},\mathrm{AB}}^\mu &= \frac{g_s^2N_c}{2}\int_k\Gamma_{\mathrm{gh}}^\alpha\Gamma_{\mathrm{gh}}^\mu\Gamma_{\mathrm{gh}}^{\alpha} D_G D_G D_Z\;.
\end{align}
This system is sufficiently simple that we make no further comment upon its solution here.

\subsection{Three-gluon vertex}
\begin{figure}[!t]
\centering\includegraphics[width=0.45\textwidth]{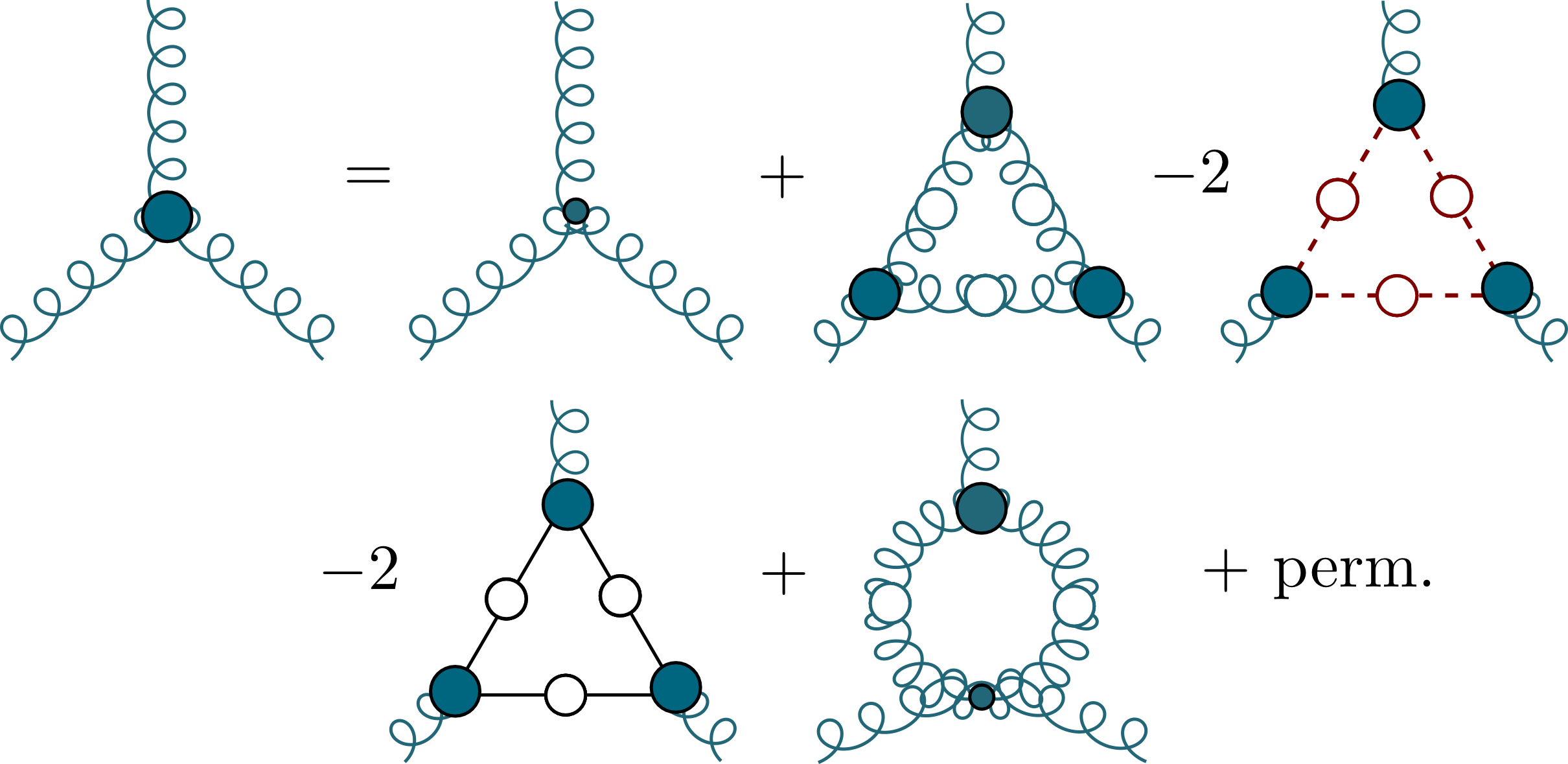}
\caption{Diagrammatic representation of the three-gluon vertex DSE. The last diagram containing 
a bare four-gluon vertex is cyclically permuted.}
\label{fig:3gvertexDSE}
\end{figure}

The DSE for the three-gluon vertex from the $3$PI effective action to three-loop, shown in 
Fig.~\ref{fig:3gvertexDSE} is
\begin{align}
\Gamma^{\mu\nu\rho}_{\mathrm{3g}} &= Z_1\Gamma^{\mu\nu\rho}_{\mathrm{3g}(0)} \nonumber \\
&+ \Lambda^{\mu\nu\rho}_{\mathrm{3g},\mathrm{GH}}
 + \Lambda^{\mu\nu\rho}_{\mathrm{3g},\mathrm{GL}}
 + \Lambda^{\mu\nu\rho}_{\mathrm{3g},\mathrm{SF}}
 + \Lambda^{\mu\nu\rho}_{\mathrm{3g},\mathrm{QL}}\;,
\end{align}
where $Z_1=Z_3/\tilde{Z}_3$ is the three-gluon vertex renormalization constant. The components 
for the ghost-loop (GH), gluon-loop (GL), swordfish (SF) and quark-loop (QL) are
\begin{align}
\Lambda^{\mu\nu\rho}_{\mathrm{3g},\mathrm{GH}} &= -N_C g_s^2 \int_q D_G D_G D_G\Gamma^\rho_{\mathrm{gh}}\Gamma^\nu_{\mathrm{gh}}\Gamma^\mu_{\mathrm{gh}}\;, \\
\Lambda^{\mu\nu\rho}_{\mathrm{3g},\mathrm{GL}} &= \frac{N_C g_s^2}{2} \int_q D_Z D_Z D_Z \Gamma^{\beta\alpha\rho}_{\mathrm{3g}}\Gamma^{\alpha\gamma\rho}_{\mathrm{3g}}\Gamma^{\gamma\beta\rho}_{\mathrm{3g}}\;,\\
\Lambda^{\mu\nu\rho}_{\mathrm{3g},\mathrm{SF}} &= -\left(3\right)\frac{3N_C g_s^2}{4} \int_q D_Z D_Z\Gamma^{\beta\alpha\rho}_{\mathrm{3g}}\Gamma^{\mu\nu\beta\alpha}_{\mathrm{4g}(0)}\;, \\
\Lambda^{\mu\nu\rho}_{\mathrm{3g},\mathrm{QL}} &= -\frac{g_s^2}{2} \sum_i \int_q \mathrm{Tr}\left[\Gamma^\rho_{\mathrm{qg}}S\Gamma^\nu_{\mathrm{qg}}S\Gamma^\mu_{\mathrm{qg}}S\right]\;.
\end{align}
The prefactors are the combination of symmetry factors ($1/2$ for the swordfish and $2$ each for 
the ghost and quark loop) and color factors. The $(3)$ for the swordfish denotes that there are 
three distinct permutations of the diagram that must be considered. Note also that the quark-loop 
contribution must be summed over each quark-flavor.

Following Ref.~\cite{Eichmann:2014xya} where it was shown that the dominant tensor structure is 
the tree-level one, we use a reduced basis to describe the dressed three-gluon vertex
\begin{align}
\Gamma^{\mu\nu\rho}_{\mathrm{3g}}(p_1,p_2,p_3) = F_1 T^{\mu\alpha}_{(p_1)}T^{\nu\beta}_{(p_2)}T^{\rho\gamma}_{(p_3)}\Gamma^{\alpha\beta\gamma}_{\mathrm{3g}(0)}(p_1,p_2,p_3)\;,
\end{align}
with $F_1 = F_1(p_1^2,p_2^2,p_3^2)$. We can furthermore arrange $p_1^2,p_2^2,p_3^2$ into a set of 
$S_3$ permutation group variables, see Appendix~\ref{sec:phasespace}, and exploit the observation 
that $s_0 = \left(p_1^2+p_2^2+p_3^2\right)/6$ is the dominant variable. Finally,
\begin{align}
\Gamma^{\alpha\beta\gamma}_{\mathrm{3g}(0)}(p_1,p_2,p_3) &= \\
 \delta^{\alpha\beta}\left(p_1 - p_2\right)^\gamma
&+\delta^{\beta\gamma}\left(p_2 - p_3\right)^\alpha
+\delta^{\gamma\alpha}\left(p_3 - p_1\right)^\beta\;,\nonumber
\end{align}
is the tree-level tensor structure of the three-gluon vertex.

\section{Results}\label{sec:results}
All our calculations for the three-point functions are performed with full momentum dependencies. 
For the presentation of results, however, we concentrate on the soft-gluon limit in which one 
gluon momentum is vanishing, $p_3=0$, whilst the remaining two legs carry the same momentum, 
$p_1=p_2=p$. This enables us to easily compare with existing and future lattice calculations. In 
terms of $S_3$ permutation group variables, this corresponds to the top of the Mandelstam 
triangle, see Fig.~\ref{fig:mandelstamplane} in appendix \ref{sec:phasespace}, with $s_0= p^2/3$, 
$a=0$, $s=1$. For the three-gluon vertex, where the dressing function is near-independent of all 
but the variable $s_0= \left( p_1^2+p_2^2+p_3^2 \right)/6$, one may make use of $s_0$ and $p$ 
interchangeably up to the obvious rescaling; this observation could be used to combine multiple 
phase space slices obtained on the lattice.

\subsection{Quark Propagator}
\begin{figure}[!t]
\centering\includegraphics[width=0.45\textwidth]{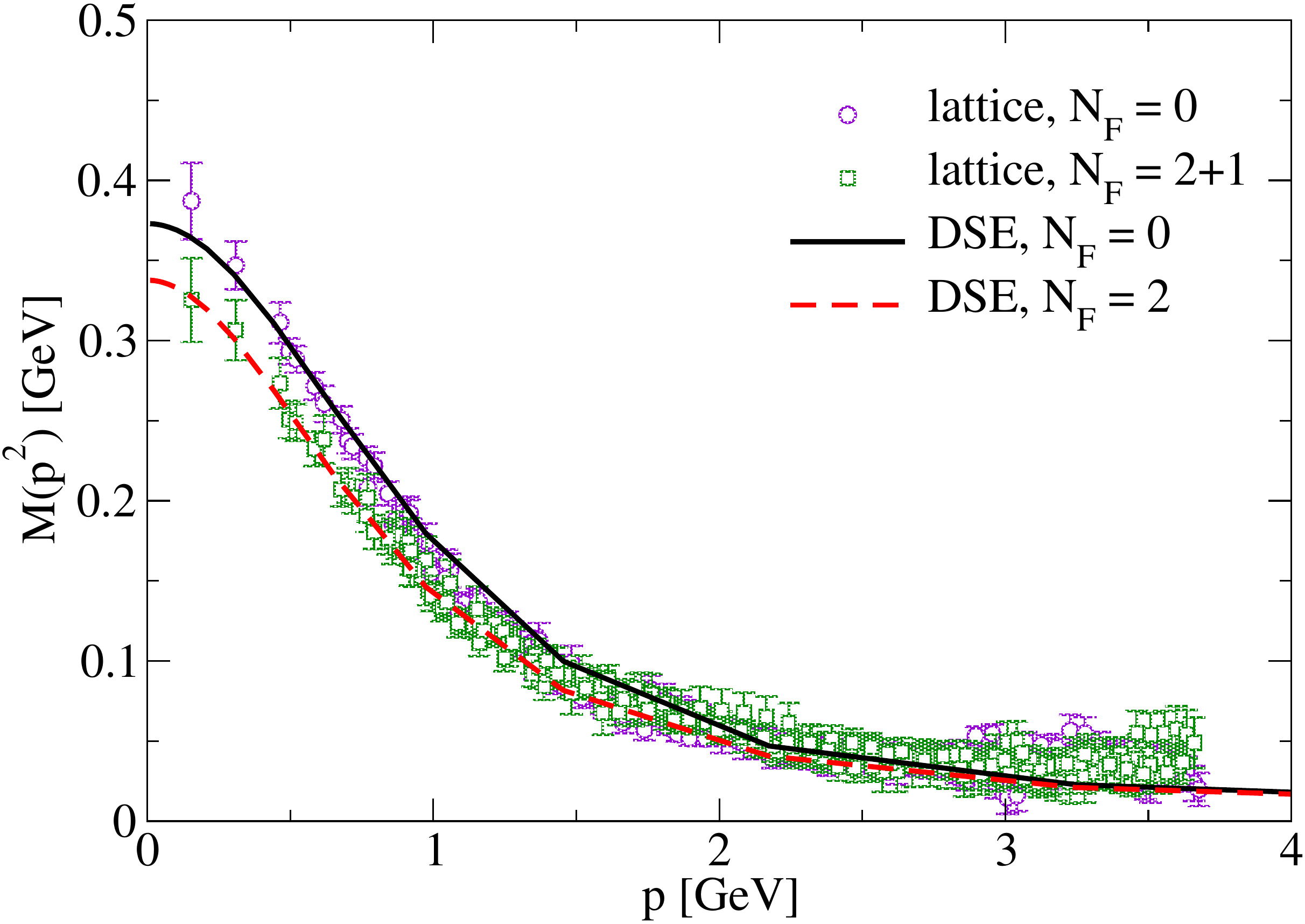}
\centering\includegraphics[width=0.45\textwidth]{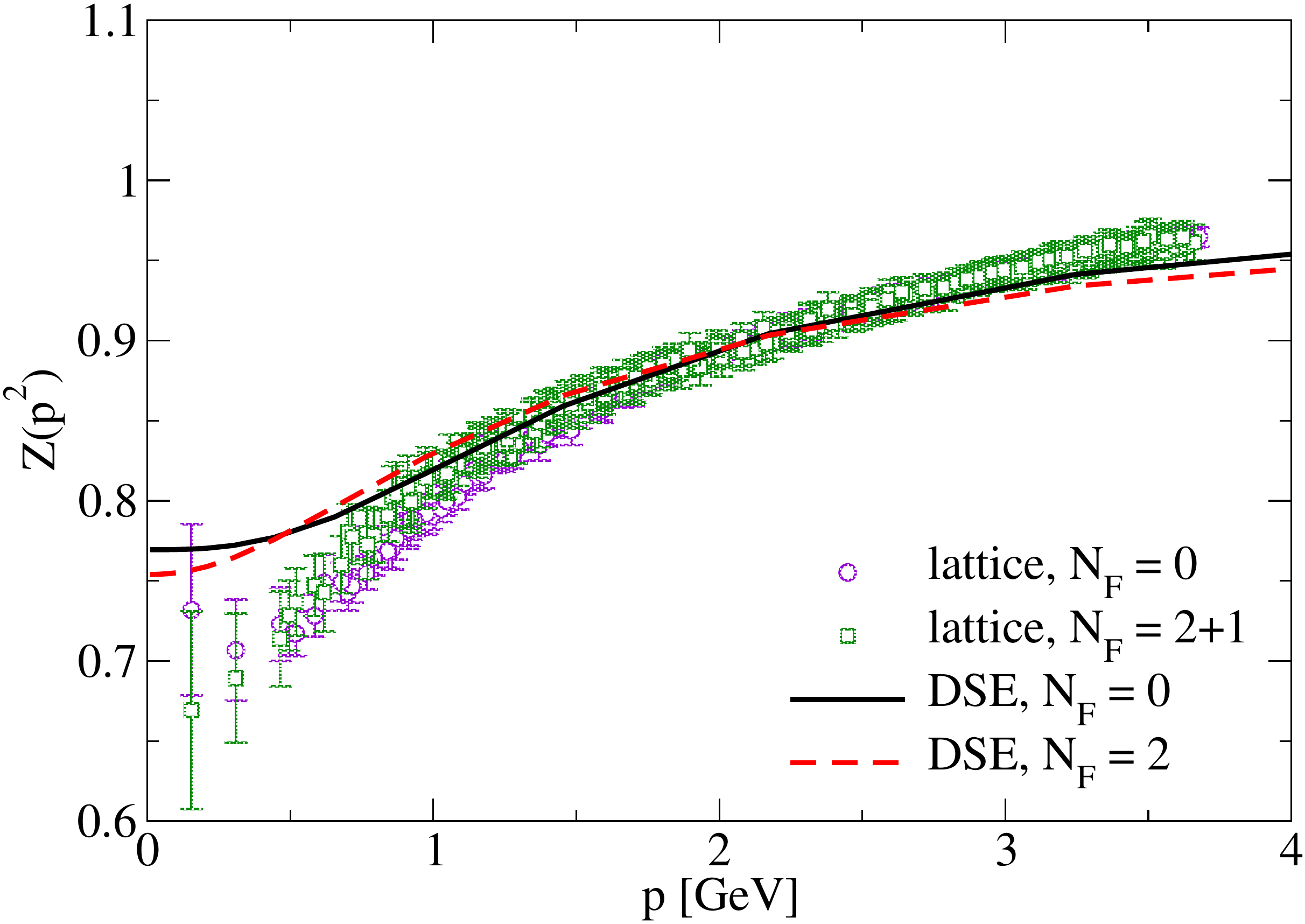}
\caption{The (top) quark mass function and (bottom) quark wave function calculated
from the 3PI effective action, compared to quenched and unquenched Lattice
data~\cite{Bowman:2005vx}.\label{fig:quarkZMlattice}}

\end{figure}

In Fig.~\ref{fig:quarkZMlattice} we show the quark mass function and wave function obtained from 
the three-loop truncation of the $3$PI effective action. We show DSE results for the quenched and 
the unquenched case with $N_f=2$. In the latter case quark loop effects in the ghost and gluon 
propagators as well as in the vertex-DSEs are taken into account. The DSE results are compared to 
quenched and unquenched lattice calculations~\cite{Bowman:2005vx}. Note the unquenched lattice 
data is for $2+1$ flavors. However, since the majority of the unquenching effects stems from the 
two light quarks, the comparison is still meaningful. Within error bars, we find very good 
agreement between our results and the lattice data, with about the right size of unquenching 
effect seen in the mass function and still room for the inclusion of a third quark flavor as well 
as for additional unquenching effects. These may evolve e.g. from the inclusion of a pion 
cloud~\cite{Fischer:2008sp} and only become apparent at higher order than the $3$PI truncation 
considered here. Note also that the wave function of the quenched propagator crosses that of the 
unquenched one, around $p=0.4$~GeV. This is a feature suggested by the lattice data that has not 
been seen in a DSE study before. We attribute this feature to the interplay of the different 
tensor structures of the fully dressed quark-gluon vertex that is only accessible in a 
self-consistent diagrammatic calculation.

\subsection{Quark-gluon vertex}
In the following we report only results for the quark-gluon vertex upon neglecting the
Abelian contribution; this is to provide consistency with the application to bound-states
considered later, see section~\ref{sec:BS}. We explicitly checked, however, that the
inclusion of the Abelian contribution has very little impact on the system of propagators
and vertices (on the level of few percent).

In the soft-gluon limit the quark gluon vertex reduces (after adapting to our conventions), to the 
form
\begin{align}
\Gamma^\mu_{\mathrm{qg}}(p,p,0)=\lambda_1\; \gamma^\mu - 4\lambda_2\;\slashed{p}p^\mu + 2\ii \lambda_3\;p^\mu\;,
\end{align}
where $\lambda_i=\lambda_i(p,p,0)$. This is the usual Ball-Chiu construction of the 
vertex~\cite{Ball:1980ay}. These dressing functions are related to those of our vertex in 
\eqref{eqn:transversebasis} by $\lambda_1 = h_1$,  $\lambda_2 = -h_2/4$, and $\lambda_3 = h_3/2$.

\begin{figure}[!b]
\centering\includegraphics[width=0.45\textwidth]{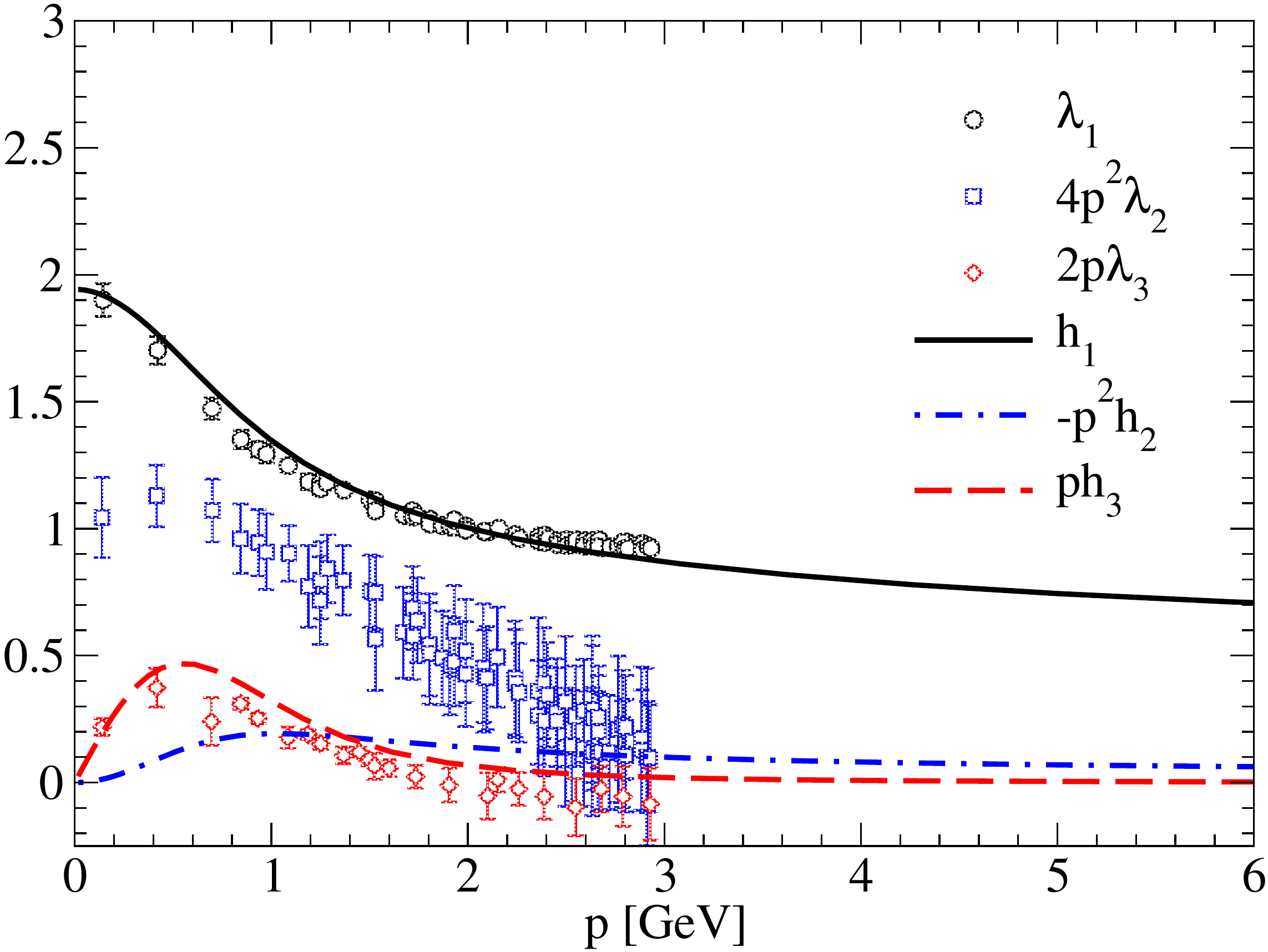}
\centering\includegraphics[width=0.45\textwidth]{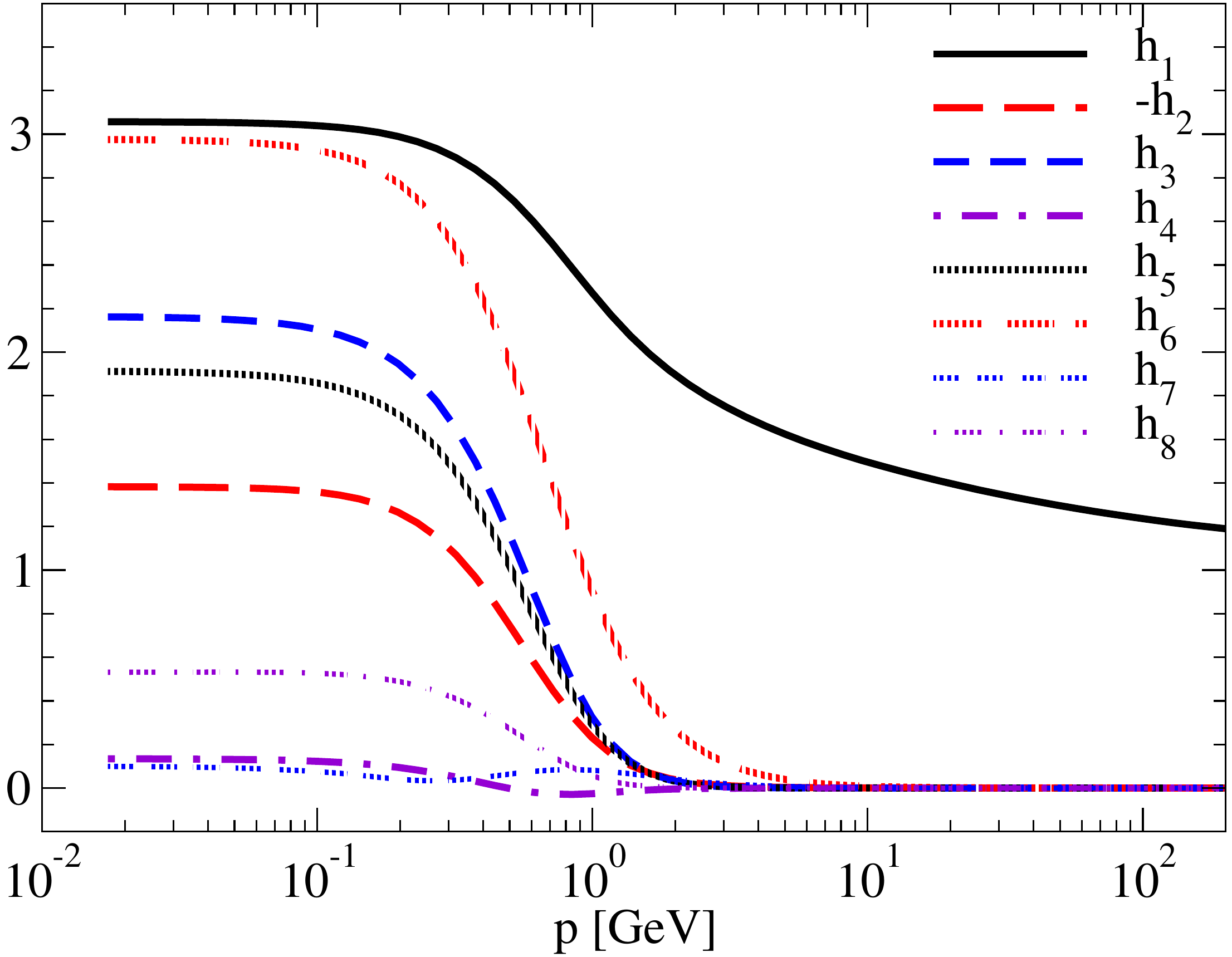}
\caption{(top) Calculated quark-gluon vertex compared to lattice calculations~\cite{Skullerud:2003qu} in the soft-gluon limit for quenched QCD; (bottom) The components of our 
unquenched quark-gluon vertex, also in the soft-gluon limit (the quenched results are 
similar).\label{fig:vertexqgquenchedunquenched}}
\end{figure}

In the top of Fig.~\ref{fig:vertexqgquenchedunquenched} we show the result of our calculation of 
the quenched quark-gluon vertex, transformed to the Ball-Chiu basis and compared with the lattice 
data of Ref.~\cite{Skullerud:2003qu}. While the $\lambda_1$ and $\lambda_3$ components are 
comparable (we introduced a vertical multiplicative shift in $\lambda_1$ to account for 
differences in the renormalization
scheme), no agreement is seen in the $\lambda_2$ terms. This component is notoriously difficult to 
extract on the lattice (the product $4p^2\lambda_2$ should vanish at the origin) and consequently 
obtained large error bars. It remains to be seen whether future more precise lattice calculations 
still retain this discrepancy. In the bottom of Fig.~\ref{fig:vertexqgquenchedunquenched} we show 
our result for the unquenched quark-gluon vertex for the basis employed 
in~\eqref{eqn:transversebasis}; for the purposes of plotting, we reversed the sign of the $h_2$ 
component. At present there are no available lattice data for the unquenched quark-gluon vertex 
with which to make a comparison. However, our results are not dissimilar to those reported in 
Ref.~\cite{Williams:2014iea} which corresponds to a three-loop expansion of the 2PI effective 
action. It would be interesting to compare with other truncations of the quark-gluon vertex in the 
DSE approach~\cite{Hopfer:PhD,Windisch:PhD} and from the functional renormalization 
group~\cite{Mitter:2014wpa}. However, the results therein have not yet been reported in a form 
that enables us to easily make a direct comparison. This is therefore relegated to future work.

\subsection{Ghost-gluon vertex}
\begin{figure}[!b]
\centering\includegraphics[width=0.45\textwidth]{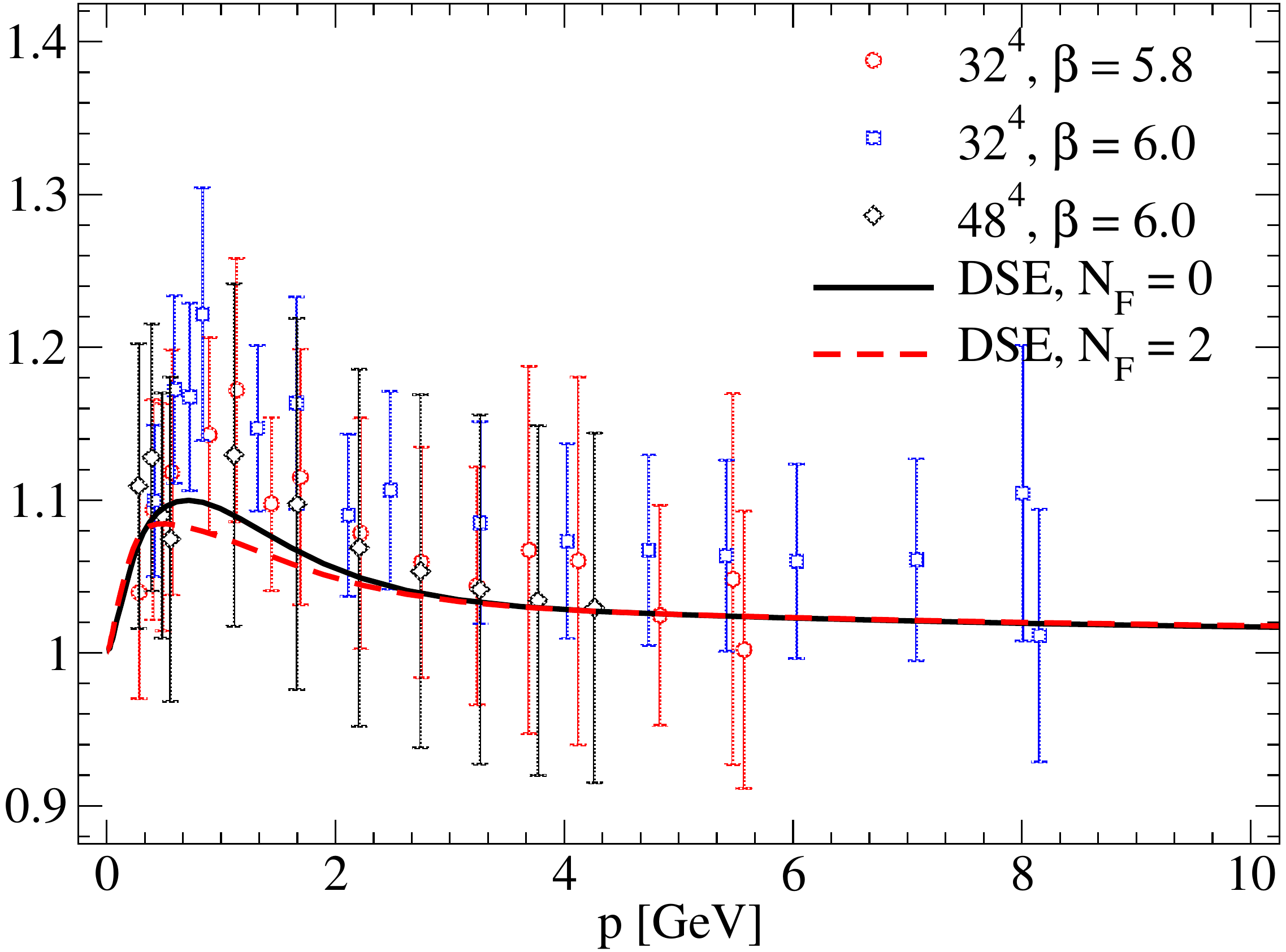}
\caption{The ghost-gluon vertex in the soft-gluon limit, as calculated from its DSE in 
\eqref{fig:ghvertexDSE}. The data points correspond to lattice data from 
Refs.~\cite{Ilgenfritz:2006he,Sternbeck:2006rd}.\label{fig:ghvertex}}
\end{figure}
For completeness we determine the dressing of the ghost-gluon vertex, in spite of the deviation of 
its tree-level term being small in Landau gauge. The result, for both the quenched and unquenched 
systems, is shown in Fig.~\ref{fig:ghvertex} and compared to extant lattice data. There is good 
agreement between our determination and that of the larger $48^4$ 
lattice~\cite{Ilgenfritz:2006he,Sternbeck:2006rd}, although clearly more work needs to be done to 
reduce the statistical errors. The impact of unquenching on the system is somewhat negligible; 
thus the oft-employed approximation that the ghost-gluon vertex can be taken bare remains a good 
one. Compared with previous DSE studies~\cite{Huber:2012kd,Aguilar:2013xqa}, the deviation of our 
ghost-gluon vertex from one is a factor of 2 or 3 smaller, which is a result of the three-gluon 
vertex being dressed in the $3$PI approach.

\subsection{Three-gluon vertex}\label{3gsec}
\begin{figure}[!b]
\centering\includegraphics[width=0.45\textwidth]{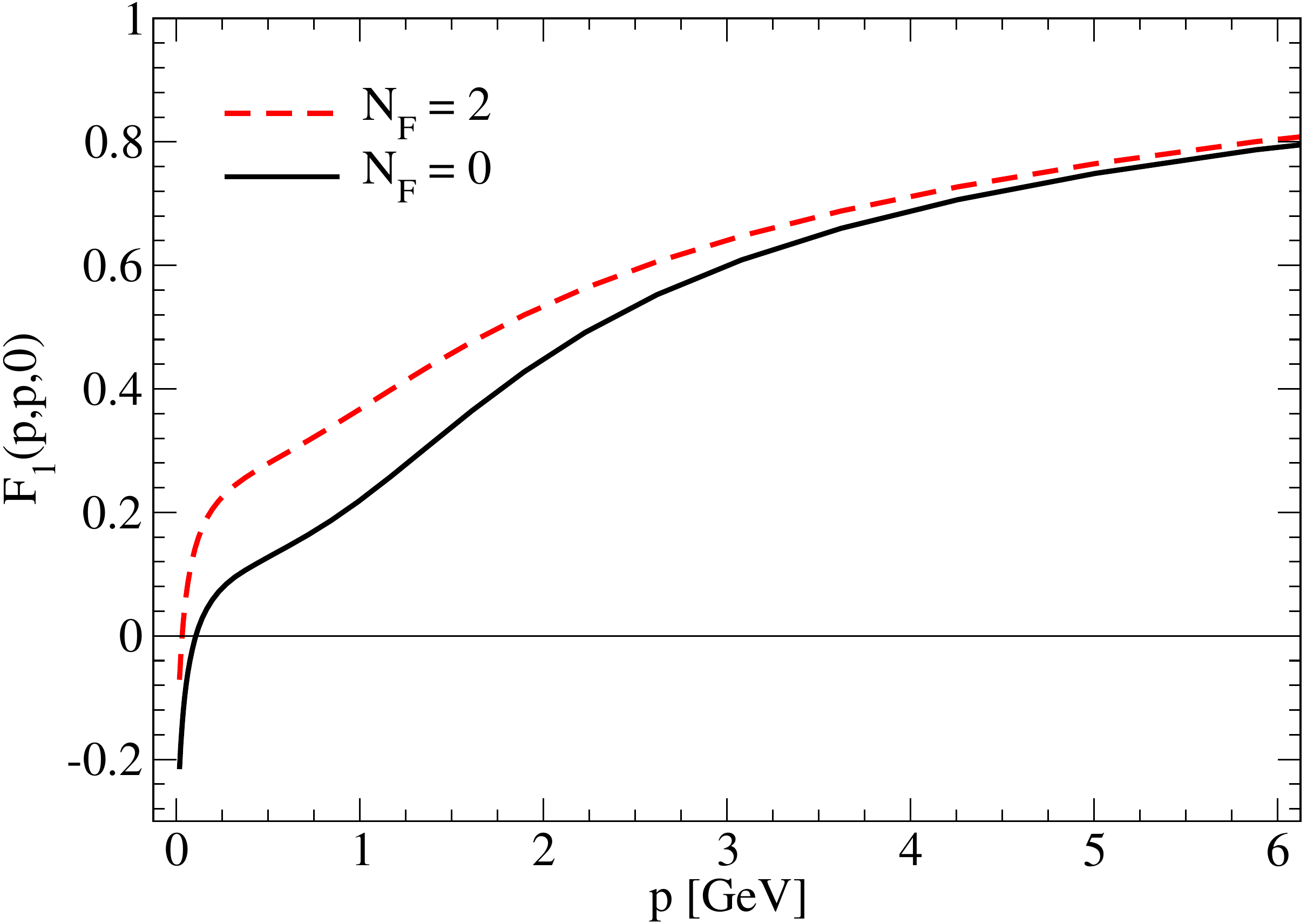}
\centering\includegraphics[width=0.45\textwidth]{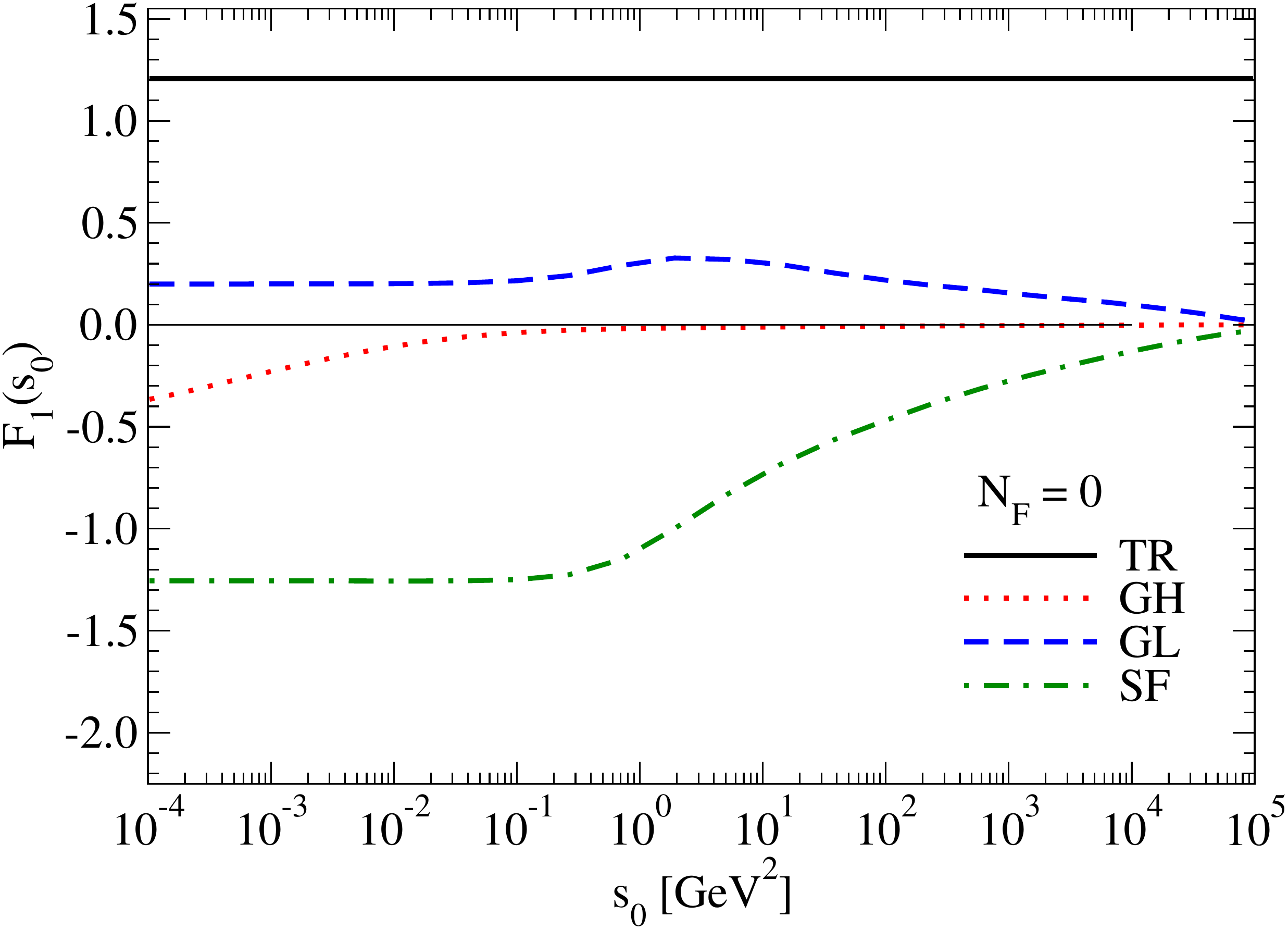}
\centering\includegraphics[width=0.45\textwidth]{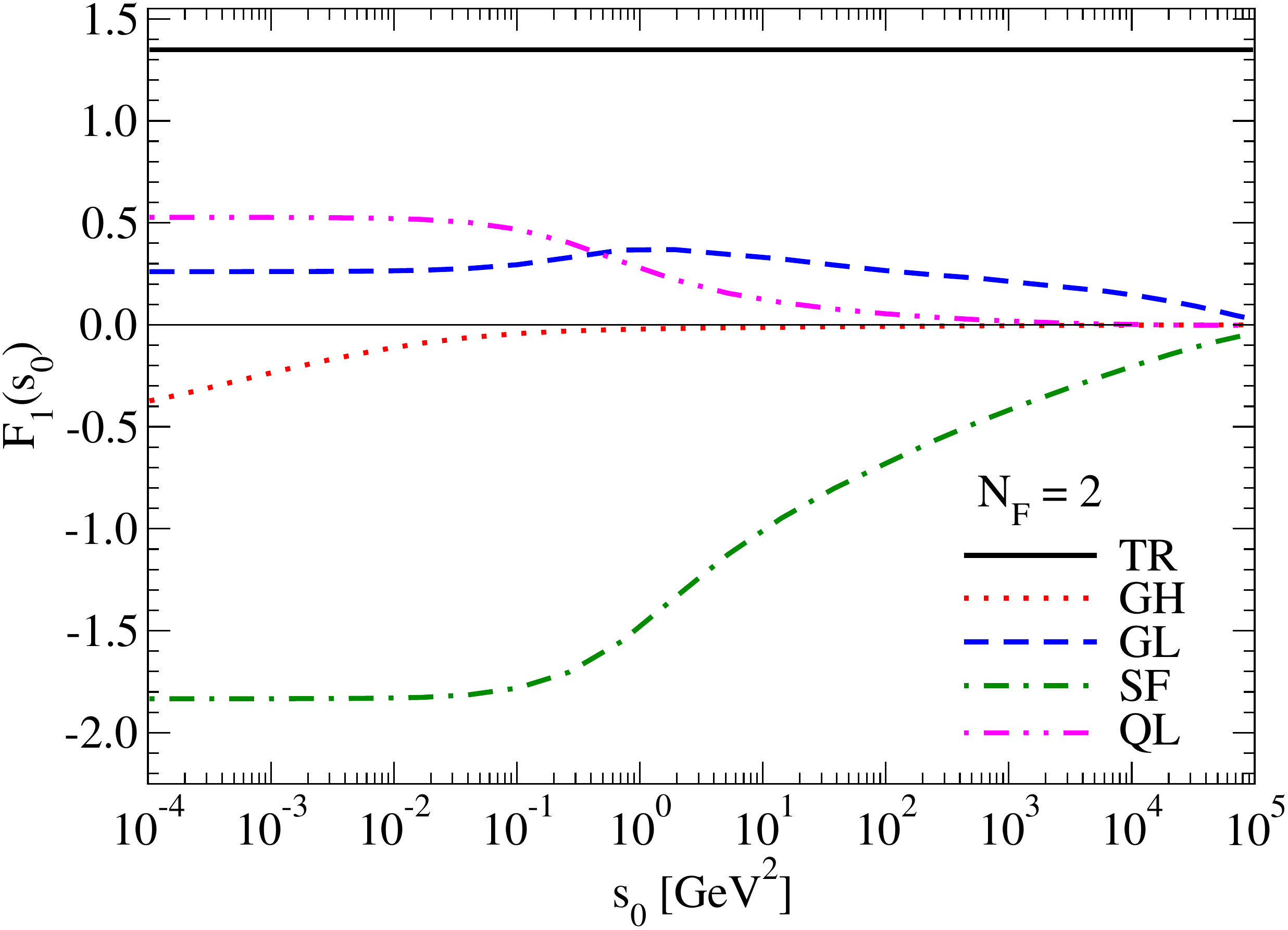}
\caption{(top) The three-gluon vertex, in the soft-gluon limit, for quenched vs unquenched QCD; 
Components contributing to the three-gluon vertex for quenched (middle) and unquenched (bottom) 
QCD. In the legend, we use abbreviations for
tree-level (TR), ghost-loop (GH), gluon-loop (GL), swordfish (SF) and quark-loop (QL).}
\label{fig:vertex3gdressing}
\end{figure}

The leading component of the three-gluon vertex, for both quenched and unquenched configurations, 
is shown in the top panel of Fig.~\ref{fig:vertex3gdressing}. In both cases, there is a zero 
crossing present below $p=0.2$~GeV in the soft-gluon kinematics. This is too low to have a 
discernible impact on the properties of hadrons. Nevertheless, the strong running of the vertex 
from its ultraviolet perturbative momentum dependence down to values close to zero in the infrared 
clearly shows that the dynamics of this vertex is an important ingredient in any calculations and 
may not be neglected. Curiously, while the impact of quark-loops on the gluon propagator is a 
reduction in its strength (see Fig.~\ref{sec:ghostgluon}), the opposite appears true in the 
three-gluon vertex where unquenching effects are clearly additive\footnote{Note that this is in 
contradiction to the preliminary results reported in Ref.~\cite{Blum:2015lsa}. We attribute this 
to a potential global sign error in their quark loop contribution. We checked explicitly that our 
sign leads to the correct flavor dependence of the anomalous dimension of the 
vertex~\cite{Pascual:1984zb}.}, see the lower panel of Fig.~\ref{fig:vertex3gdressing}.

In general, gluon propagators and three-gluon vertices appear in combinations on the right hand 
sides of the DSEs for the three-gluon and the quark-gluon vertex. Thus the unquenching effects in 
the gluon propagator compete against their sister contributions in the three-gluon vertex. Within 
the three-gluon vertex DSE we find that unquenching in the three-gluon vertex wins -- at least for 
two quark flavors -- and consequently the unquenched three-gluon vertex is less suppressed, as 
compared to its tree-level value, than the quenched system. For this reason, unquenching effects 
in the gluon propagator and the three-gluon vertex partly balance each other also in the DSE for 
the quark-gluon vertex. Thus, the net unquenching effect on the quark-gluon vertex and quark mass 
function is not as dramatic as one may expect from the gluon propagator alone.

\subsection{Application to bound-states}\label{sec:BS}
Before we present results for the bound-states, it is appropriate to discuss some of the 
implications of choosing the top-down approach wherein the emphasis is upon reproducing the 
Green's functions of QCD (i.e. those that are in agreement with lattice calculations). Since the 
system we consider is ultimately a truncation of the effective action it is not guaranteed that 
the resulting meson spectrum will be in agreement with experiment. That is, the scales inherited 
from the Lattice ghost and gluon propagators may not yield a phenomenologically precise value e.g. 
for the pion decay constant. Nonetheless, as the first work that incorporates a self-consistently 
solved 3PI-system including bound states, we choose the top-down approach and relegate a thorough 
discussion of the realisation of symmetry constraints and their impact on the low-lying 
meson/baryon spectrum to a future paper.

\begin{figure}[!t]
\centering\includegraphics[width=0.45\textwidth]{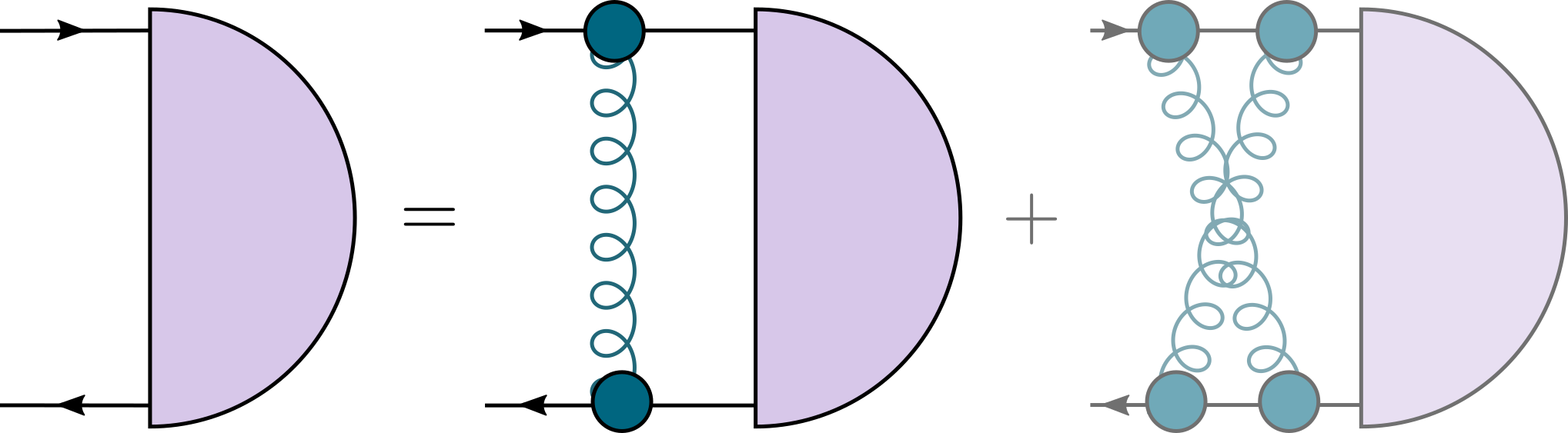}
\caption{Symmetry preserving Bethe-Salpeter equation corresponding to the truncation of the 
quark-gluon vertex in Fig.~\ref{fig:qgvertexDSE}. The crossed ladder term, stemming from the 
abelian diagram in the vertex DSE, is displayed but discarded for the calculation of the 
bound-states.}
\label{fig:bsekernel}
\end{figure}

The Bethe-Salpeter kernel
corresponding to the truncation at hand can readily be derived from the $3$PI effective 
action~\cite{Cornwall:1974vz,Fukuda:1987su,Komachiya:1989kc,McKay:1989rk,Carrington:2010qq,Sanchis-Alepuz:2015tha}. 
In the case of the Bethe-Salpeter equation (BSE) for a meson, its quark-antiquark kernel is 
obtained by twice differentiating with respect to the quark propagator $S$; in 
Appendix~\ref{sec:AXWTI} we discuss how consistency with the axial-vector Ward-Takahashi identity 
leads to the appearance of a Goldstone boson in the chiral limit and the formation of the 
Gell-Mann-Oakes-Renner relation.
The result, following simplification upon imposing the stationary condition, is given in 
Fig.~\ref{fig:bsekernel} and features a one-gluon exchange contribution and a crossed-ladder 
exchange, with all propagators and vertices fully-dressed. Though the crossed-ladder diagram 
is $N_c^2$ suppressed with respect to the leading gluon exchange (similar to the Abelian vs. 
non-Abelian diagram in the quark-gluon vertex), it may yet be relevant due to the dynamical 
enhancement contained within the vertices. However, for reasons of expediency (i.e. to avoid a 
complicated two-loop term in the BS equation) we will not include the crossed-ladder term here. 
This can easily be made consistent at the level of the effective action by also dropping the 
corresponding diagram.

It follows that the Bethe-Salpeter equation is
\begin{align}
\Gamma = C_F g_s^2\int_k\Gamma^\mu_{\mathrm{qg}}S \Gamma S  \Gamma^\mu_{\mathrm{qg}}D_Z\;,
\end{align}
where once again we have omitted momentum arguments for brevity.  The quantum numbers of the 
amplitude under consideration, $\Gamma(p,P)$, are dictated by its tensor 
decomposition~\cite{LlewellynSmith:1969az,Krassnigg:2010mh,Fischer:2014xha}.
In a compact notation, we can solve this homogeneous equation
\begin{align}\label{eqn:homogeneousBSE}
\Gamma_i = \lambda(P^2) K_{ij}\Gamma_j\;,
\end{align}
as an eigenvalue equation, where $\lambda(P^2)=1$ gives solutions at discrete values of the 
bound-state total momentum squared, $P^2=-M_i^2$. The matrix $K_{ij}$ represents the integral 
kernel in the BSE.

As noted previously, access to time-like properties of bound-states (such as their masses), 
requires an analytic continuation to complex momenta. By carefully choosing the momentum routing, 
we can arrange the coupled system of DSEs to be such that only the quark and quark-gluon vertex 
need to be analytically continued to the complex plane. The subsequent evaluation of the quark and 
quark-gluon vertex is accomplished using a combination of the Cauchy contour 
method~\cite{Fischer:2005en,Krassnigg:2009gd}, and the shell method~\cite{Fischer:2008sp}. 
Combining these two techniques together proves to be not only reliable, but also efficient.

\begin{table}[!t]
\centering
\caption{Meson masses and pion decay constant in GeV as calculated in rainbow-ladder 
(RL)~\cite{Maris:1999nt}, the 2PI effective action at 3-loop 
($2$PI-3L)~\cite{Sanchis-Alepuz:2015qra} and in the 3PI effective action at 3-loop ($3$PI-3L) 
truncation as detailed here\label{tab:mesons}, compared to values from the Particle Data Group 
(PDG)~\cite{Agashe:2014kda}. Results affixed with ${}^\dag$ are fitted values.}
\setlength{\tabcolsep}{1em}
\begin{tabular}{|c||c|c|c|c|}
\hline
\hline
                             &    RL          &       $2$PI-3L  &      $3$PI-3L         &     PDG \\
\hline
\hline
       $0^{-+}~(\pi)$        &   $0.14^\dag$  &     $0.14^\dag$ &      $0.14^\dag$      &   $0.14\phantom{(0)}$    \\
       $0^{++}~(\sigma)$     &   $0.64$       &     $0.52$      &      $1.1(1)$         &   $0.48(8)$ \\
       $1^{--}~(\rho)$       &   $0.74$       &     $0.77$      &      $0.74$           &   $0.78\phantom{(0)}$    \\
       $1^{++}~(a_1)$        &   $0.97$       &     $0.96$      &      $1.3(1)$         &   $1.23(4)$ \\
       $1^{+-}~(b_1)$        &   $0.85$       &     $1.1$       &      $1.3(1)$         &   $1.23\phantom{(0)}$    \\
\hline
       $f_\pi$               &   $0.092^\dag$ &   $0.103$       &      $0.105$          &   $0.092$                \\
\hline
\hline
\end{tabular}
\end{table}

The results of our calculation in the $3$PI effective action at 3-loop ($3$PI-3L) are detailed in 
Table~\ref{tab:mesons} and contrasted with typical results from rainbow-ladder 
(RL)~\cite{Maris:1999nt}, a recent study of mesons (and baryons) in the framework of the $2$PI 
effective action at 3-loop ($2$PI-3L)~\cite{Sanchis-Alepuz:2015qra}, and of course their 
experimentally known values from the PDG~\cite{Agashe:2014kda}. We use the unquenched $N_F=2$ 
system throughout.

We see that the pion appears as a pseudo-Goldstone boson in all truncations; its mass is fitted to 
$140$~MeV in each case so as to determine the light quark mass used as input. As has been 
previously noted, that the system exhibits the correct chiral dynamics ensures that the mass of 
the vector meson is reproduced on the level of $5\%$.

One of the problems with the rainbow-ladder approach is its inability to reproduce the correct 
splitting of the axial-vectors and the $\rho$ meson; that is, the axial-vectors are typically too 
light by several hundred MeV. This was partially remedied by including three-loop corrections in 
the $2$PI effective action, lifting the size of one of the axial-vectors and thus suggesting that 
tensor structures in the quark-gluon vertex beyond the tree-level play an important role. Here, we 
find that the vector axial-vector splitting is $0.56(10)$~GeV for both charge conjugation states; 
this is of the same order as expected from experiment.

Similarly, whilst in RL and the $2$PI-3L calculations there remains a light scalar around 
$0.5$-$0.6$~GeV (without a width) that complicates the interpretation of the $f_0(500)$ as a 
four-quark state~\cite{Heupel:2014ina}, the present calculation lifts the lightest scalar to be 
$1.1(1)$~GeV in line with our expectations. These results agree with those of the bottom-up 
approach of~\cite{Chang:2011ei}, where an effective quark-gluon interaction is constructed to 
reproduce the vector axial-vector splitting with a heavy scalar. It will be interesting to see how 
our calculated top-down approach compares to the phenomenologically constrained one in detail.

Since the components of our $3$PI quark-gluon vertex are similar to those of the $2$PI-3L 
truncation, it appears that it is the difference in the structure of the kernel itself that leads 
to these improvements. Indeed, since the leading part of the kernel is a gluon-ladder that 
connects two fully-dressed quark-gluon vertices, we have for the first time included explicit 
scalar-scalar terms in addition to the usual vector-vector and, recently, vector-scalar ones 
considered thus far. It will be the topic of a future study to see how these effects impact upon 
the excited state spectrum, in particular that of charmonium, and what the consequences will be 
for baryons.

\section{Conclusion}\label{sec:conclusion}
We calculated the quark-gluon vertex in the three-loop truncation of the $3$PI effective action, 
neglecting for now the backcoupling of the calculated vertices on the underlying ghost and gluon 
propagators, which are instead fixed separately such that agreement with the Lattice is obtained. 
We find that the leading part of the vertex is strongly enhanced for light quarks, as seen in 
previous studies, and that the sub-leading components are already quite stable in comparison to, 
for example, a $2$PI truncation of the effective action to third loop order.

In addition, we investigated the impact of including unquenching effects in the form of quark-loop 
corrections to both the gluon propagators and three-gluon vertex. We find that they are sizeable 
in the latter, introducing a material shift in the location of the zero crossing to momenta deeper 
in the infrared. We also observed that the effects of unquenching in the three-gluon vertex act in 
opposition to those in the gluon propagator; as a consequence, the overall impact on the quark 
propagator and quark-gluon vertex are smaller than one would naively expect.

By neglecting the Abelian contribution to the quark-gluon vertex -- which by itself can be 
calculated without difficulty -- we were able to apply the present framework to the calculation of 
quark-antiquark bound-states via the Bethe-Salpeter equations; the kernel of which was derived 
from the $3$PI effective action and is used in accordance with the axial Ward-Takahashi identity 
to ensure the appearance of the pion as a Goldstone boson in the chiral limit. To obtain 
bound-state masses in the time-like region, we analytically continued the quark and quark-gluon 
vertices to complex momenta. In contrast to previous top-down studies we find that the lightest 
scalar is above $1$~GeV, thus adding further evidence in support of the tetraquark picture of the 
$f_0(500)$ \cite{Heupel:2012ua,Eichmann:2015cra}, as well as reproducing the correct mass 
splitting between vector and axial-vector states.

To improve upon the present work, we would have to include solutions of the ghost and gluon 
propagators obtained self-consistently from the $3$PI effective action. This would require that 
the inherent two-loop gluon polarizations be included. Additionally, the Abelian contribution to 
the quark-gluon vertex should be included with the difficult crossed-ladder term incorporated into 
the Bethe-Salpeter kernel.  All of these tasks pose sizeable challenges -- towards which we have 
made significant progress -- and are thus relegated to a future work.

\section{Acknowledgements}
We thank G.~Eichmann and H.~Sanchis-Alepuz for useful discussions and a critical reading of this 
manuscript. We are grateful to Andre Sternbeck for sending us his lattice data for the unquenched 
gluon and ghost propagators prior to publication. This work has been supported by the Helmholtz 
International Center for FAIR within the LOEWE program of the State of Hesse and by BMBF under 
contract 06GI7121.

\appendix

\section{Technical Details}\label{sec:appendix}
\subsection{Ghost and gluon propagators}\label{sec:ghostgluon}

\begin{figure}[!t]
\centering\includegraphics[width=0.45\textwidth]{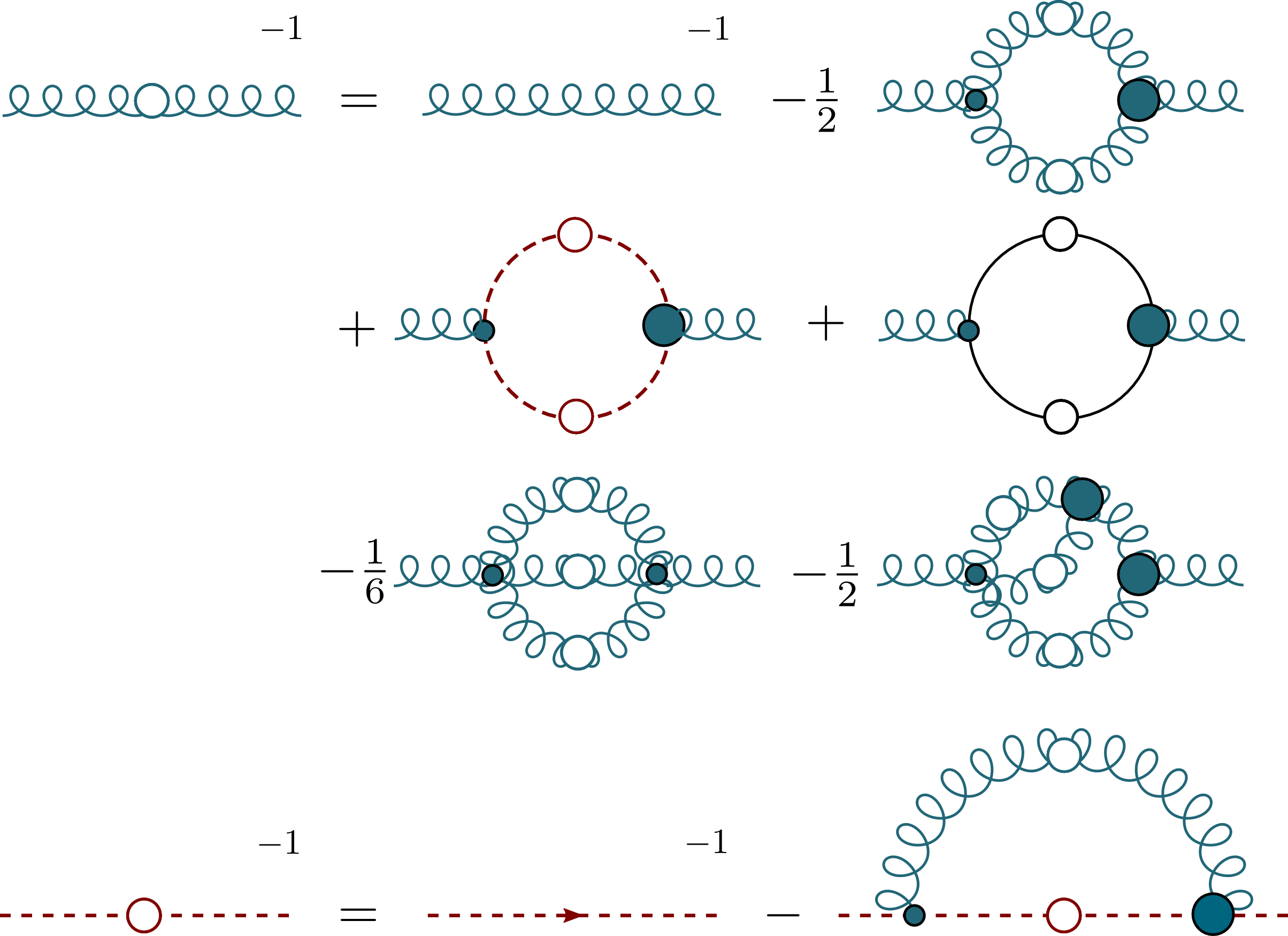}
\caption{The DSEs for the ghost and gluon propagators. In our truncation we neglect any 
contributions from the four-gluon vertex, i.e. the two-loop graphs are not included.}\label{fig:ghostgluondse}
\end{figure}

In the main text we explained the need to solve the DSEs for the quenched and unquenched ghost and 
gluon propagators using effective three-point vertices that incorporate the missing two-loop 
diagrams in the gluon DSE. The resulting system of equations is shown in 
Fig.~\ref{fig:ghostgluondse}. Here, we specify the vertex models used to solve this system of 
equations as well as the corresponding renormalization conditions. We choose 
$\alpha(\mu^2) = 0.124$ at the renormalization point $\mu = 57$ GeV together with the MiniMOM 
condition for the renormalization factor of the ghost-gluon vertex, 
$\tilde{Z}_1=1$~\cite{vonSmekal:2009ae}. Moreover, we need to single out one instance of the 
one-parameter family of decoupling solutions, see Refs.~\cite{Boucaud:2008ky,Fischer:2008uz} for 
details. This is done by imposing the condition $G(0)=3.8$ on the ghost dressing.
These coupling and renormalization constants are carried over to the whole system of DSEs that are 
solved in the main body of this work, where the calculated ghost and gluon propagators serve as 
input.

For the ghost-gluon vertex model we use the simplest possible choice: the bare vertex. This choice 
is well justified in Landau gauge, as also discussed in the main text. For the three-gluon vertex 
we use the Bose-symmetric model suggested by Huber and Smekal in Ref.~\cite{Huber:2012kd}
\begin{widetext}
\begin{align}
\Gamma_{3g}^{\mu\nu\rho}(p_1,p_2,p_3) &= Z_1^{-1}\,\Gamma_{3g, 0}^{\mu\nu\rho} \,D^{A^3,UV}(p_1,p_2,p_3) \,D^{A^3}(p_1,p_2,p_3)\;,\\
D^{A^3,UV}(p_1,p_2,p_3)               &= \left[G\left(3s_0\right)\right]^{3+1/\delta}\;,\\
D^{A^3}(p_1,p_2,p_3)                  &= D^{A^3,UV}(p_1,p_2,p_3) + D^{A^3,IR}(p_1,p_2,p_3)\;,\\
D^{A^3,IR}(p_1,p_2,p_3)			      &= h_{IR}\,\left[G(6s_0)\right]^3\, \left[f^{3g}(p_1^2) f^{3g}(p_2^2) f^{3g}(p_3^2)\right]^4\;.
\end{align}
\end{widetext}
Here $s_0 = \left(p_1^2+p_2^2+p_3^2\right)/6$ parametrizes the leading scale dependence, and 
$f^{3g}(x)=\left(1 + x/\Lambda^2_{3g}\right)^{-1}$ is an infrared damping factor. The tensor 
structure of the bare three-gluon vertex is $\Gamma_{3g, 0}^{\mu\nu\rho}$, the vertex 
renormalization factor is $Z_1$, and $\delta=-9 N_c/(44 N_c - 8 N_f)$ is the anomalous dimension 
of the ghost. The contribution $D^{A^3,UV}$ ensures the correct ultraviolet running of the vertex 
and $D^{A^3,IR}$ its damping in the infrared in agreement with the lattice data of the vertex and 
the numerical results, cf. \cite{Cucchieri:2008qm,Blum:2014gna,Eichmann:2014xya} and our results 
in the main body of this work. The damping is controlled by two parameters, $h_{IR}$, 
$\Lambda_{3g}$, for which we choose $h_{IR}=-1$, $\Lambda_{3g}=1.3$~GeV in the quenched case and 
in accordance with \cite{Huber:2012kd}. For the unquenched calculation ($N_f=2$) we had to modify 
these values to $h_{IR}=-0.1$, $\Lambda_{3g}=3.6$ GeV. Interestingly, these changes are 
qualitatively in agreement with the corresponding changes in our numerical results for the 
three-gluon vertex from its 3PI-DSE, discussed in section \ref{3gsec}.

For the quark-gluon vertex we employ a vertex ansatz that has been introduced in 
Ref.~\cite{Fischer:2003rp,Fischer:2005en} and is constructed along the (leading part) of the 
Slavnov-Taylor identity for the vertex. With quark momenta $p_1$ and $p_2$ and gluon momentum $p_3$
it reads:
\begin{align}
\Gamma^\mu_{\mathrm{qg}}(p_1,p_2,p_3) &= \gamma^\mu\, A(p_3^2)\, G^2(p_3^2)\, \tilde{Z}_3
\frac{\left(G(p_3^2) \,\tilde{Z}_3\right)^{-2d-d/\delta}}{\left(Z(p_3^2) \,Z_3\right)^d}\,.
\end{align}
The construction is such that the ultraviolet momentum running of the vertex is the same for all 
values of the parameter $d$, leading to the correct ultraviolet running of the quark propagator in 
agreement with resummed perturbation theory. For our two-flavor calculations we choose 
$d=1$\footnote{Note that previous unquenched calculations of ghost and gluon propagators with the 
so-called scaling infrared behaviour resulted in values of $d$ around $d=0$~\cite{Fischer:2003rp}. 
With decoupling, as adopted here, this value changes substantially.}. It has been noted already in 
Ref.~\cite{Fischer:2003rp,Fischer:2005en} that such a vertex model needs to have different 
momentum assignments when employed in the quark DSE and in the quark-loop of the gluon DSE. This 
can be shown strictly using multiplicative renormalisability. Thus we use
\begin{align}
\Gamma^\mu_{\mathrm{qg}}(p_1,p_2,p_3) &= \gamma^\mu\, \frac{A(p_1^2)+A(p_2^2)}{2}\, G(p_1^2)\,G(p_2^2)\, \nonumber\\
&\times \tilde{Z}_3
\frac{\left(G(p_1^2)G(p_2^2)\right)^{-d-d/(2\delta)}\,\tilde{Z}_3^{-2d-d/\delta}}{\left(Z(p_1^2)\,Z(p_2^2)\right)^{d/2} Z_3^d}\,,
\end{align}
in the quark loop which leads to the correct running of the ghost and gluon propagators in the 
ultraviolet momentum region.

Finally we need to specify the bare quark masses for the two light quarks. Here we chose two 
chiral quarks for simplicity. We explicitly checked that only very tiny changes result for the 
ghost and gluon propagators when these values are modified to ones in the physical  range, i.e. 
those that lead to the experimental pion mass.

\subsection{Phase space and vertices: $S_3$ permutation group}\label{sec:phasespace}
\begin{figure}[!h]
\centering\includegraphics[width=0.31\textwidth]{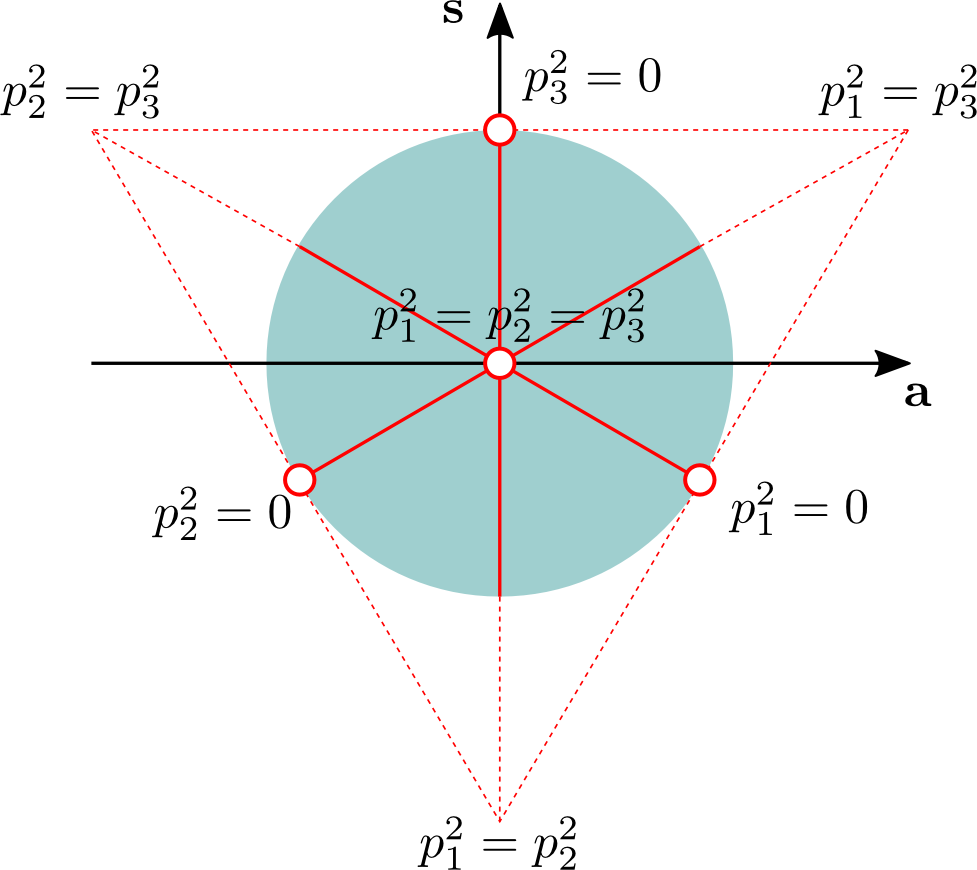}
\caption{Permutation group variables arranged into a Mandelstam plane.}
\label{fig:mandelstamplane}
\end{figure}
Useful in the study of three-point functions is the expression of momenta $p_1$, $p_2$ and $p_3$ 
in accord to the $S_3$ permutation group. Whilst this is of direct relevance for the three-gluon 
vertex~\cite{Eichmann:2014xya}, owing to bose-symmetry, it remains a useful representation also 
for both the ghost-gluon and quark-gluon vertices. In all cases, we arrange the momenta into the 
variables
\begin{align}
s_0 &= \frac{p_1^2+p_2^2+p_3^2}{6}\;, \\
a   &= \frac{\sqrt{3}\left(p_2^2-p_1^2\right)}{6s_0}\;,\\
s   &=  \frac{p_1^2+p_2^2-2p_3^2}{6s_0}\;.
\end{align}
It is straightforward to see that the doublet $(a,s)$ forms the inside of a circle, see 
Fig.~\ref{fig:mandelstamplane}, whilst the singlet $s_0$ carries an overall momentum scale. For 
actual calculations, instead of $(a,s)$ a more natural parametrization of the circle in the polar 
coordinates $(r,\psi)$ is chosen.

Consequently, symmetry properties of the vertex (provided a suitable basis is constructed) are 
reflected in the phase space variables. For example, bose-symmetry of the three-gluon vertex 
manifests as a $2\pi/3$ periodicity in the angular variables $\psi$. For the quark-gluon vertex 
(ghost-gluon vertex), charge conjugation (bose-symmetry in the ghost legs) manifests as reflection 
symmetry in the $s$-axis.

\subsection{Axial Ward-Takahashi identity}\label{sec:AXWTI}
In order for the system to feature a massless pion in the chiral limit and in the presence of 
dynamical chiral symmetry breaking, the interaction kernel of the Bethe-Salpeter equation and the 
self-energy of the quark must obey the axial-vector Ward-Takahashi identity. In the framework of 
Dyson-Schwinger and Bethe-Salpeter equations, this relation reads~\cite{Munczek:1994zz}
\begin{align}
   \frac{\delta^2 \Gamma}{\delta S \delta S}\{\gamma_5,S\} = 0\;,
\end{align}
where the curly brackets indicate an anti-commutator.
For a given effective action $\Gamma$, it is sufficient that the effective action is invariant 
under a global chiral transformation\footnote{See however 
Refs.~\cite{Pilaftsis:2013xna,Brown:2015xma} in the case where chiral symmetry is broken by 
truncation artifacts.}; for the $3$PI effective action used in this work, chiral symmetry appears 
to be invariant by inspection. What is not obvious is the mechanism  by which the invariance of 
the action is connected to the existence of a Goldstone boson. Following the steps in 
Ref.~\cite{Munczek:1994zz}, we start with the global chiral transformation properties of the quark 
$S$ and the quark-gluon vertex $\Gamma^\mu_{\mathrm{qg}}$
\begin{align}
    & S^\prime = e^{i\gamma_5\tau^l\theta}\,S\, e^{i\gamma_5\tau^l\theta}\;,
    \\\nonumber
    & \Gamma^{\mu\prime}_{\mathrm{qg}} = e^{-i\gamma_5\tau^l\theta}\,\Gamma^\mu_{\mathrm{qg}}\, e^{-i\gamma_5\tau^l\theta}\;,
\end{align}
where $\tau^l$ is a generator of the flavor group and $\theta$ is the real transformation angle. 
We checked explicitly that the quark DSE and vertex DSE are indeed invariant under such a 
\emph{combined} transformation.
Performing an infinitesimal chiral transformation of the effective action, which is zero by 
invariance, and subsequently taking a derivative with respect to the quark, the following relation 
is obtained
\begin{align}
     \frac{\delta^2\Gamma}{\delta S \delta S}\{\gamma_5,S\} -   \frac{\delta^2\Gamma}{\delta S \delta \Gamma^\mu_{\mathrm{qg}}}\{\gamma_5,\Gamma^\mu_{\mathrm{qg}}\} = 0\;,
\end{align}
where for brevity we have dropped all indices and momentum/space-time arguments. The first term is 
the sought after Goldstone boson, where the anti-commutator is the wave function of the pion and 
the second derivative of the effective action represents the BSE-operator. Thus, at first glance 
glance the second term seems to spoil the existence of a Goldstone boson. However, after a lengthy 
calculation that repeatedly employs the equation of motion of the quark-gluon vertex 
(see Fig.~\ref{fig:qgvertexDSE}) and the constraint imposed by the axial Ward identity, the 
following can be shown
\begin{align}
     \frac{\delta^2\Gamma}{\delta S \delta S}\{\gamma_5,S\} = 0\;.
\end{align}
Thus, the truncated $3$PI system features a massless pion. Further details of the systematics 
involved in this derivation will be given in a separate publication.
\bibliographystyle{apsrev4-1}
\bibliography{vertex3PI}

\end{document}